Review

## COVID-19: post infection implications in different age groups, mechanism, diagnosis, effective prevention, treatment, and recommendations

Muhammad Akmal Raheem[1,2], Muhammad Ajwad Rahim[3], Ijaz Gul[1,2], Md. Reyad-ul-Ferdous[1,2], Liyan Lei[2], Junguo Hui[2], Shuiwei Xia[2], Minjiang Chen[2], Dongmei Yu[2,4], Vijay Pandey[1], Peiwu Qin[1,2*], Jiansong Ji[2*]

[1]Institute of Biopharmaceutical and Health Engineering, Tsinghua Shenzhen International Graduate School, Tsinghua University, Shenzhen 518055, PR China.

[2]Lishui Central Hospital, Wenzhou Medical University, Lishui, P.R. China

[3]College of Animal science and Technology, Ahnui Agricultural University, Hefei, China.

[4]School of Mechanical, Electrical & Information Engineering, Shandong University.

[*]Correspondence: pwqin@sz.tsinghua.edu.cn (P.Q), ji_j_s@sina.com (J.S.J)

**Highlights**

- COVID-19 induces long-term effects in individuals of both genders of various ages.
- Artificial intelligence-based COVID-19 diagnostic tools are efficient.
- Pharmacological and non-pharmacological treatments reduced the long-term impacts of COVID-19.
- All vaccines significantly reduced the persistent effects of COVID-19.
- No vaccine provides lifetime protection against COVID-19.
- Protective measures greatly limit SARS-CoV-2 transmission.

**Abstract**

SARS-CoV-2, the highly contagious pathogen responsible for the COVID-19 pandemic, has persistent effects that begin four weeks after initial infection and last for an undetermined duration. These chronic effects are more harmful than acute ones. This review explores the long-term impact of the virus on various human organs, including the pulmonary, cardiovascular, neurological, reproductive, gastrointestinal, musculoskeletal, endocrine, and lymphoid systems, particularly in older adults. Regarding diagnosis, RT-PCR is the gold standard for detecting COVID-19, though it requires specialized equipment, skilled personnel, and considerable time to produce results. To address these limitations, artificial intelligence in imaging and microfluidics technologies offers promising alternatives for diagnosing COVID-19 efficiently. Pharmacological and non-pharmacological strategies are effective in mitigating the persistent impacts of COVID-19. These strategies enhance immunity in post-COVID-19 patients by reducing cytokine release syndrome, improving T cell response, and increasing the circulation of activated natural killer and CD8 T cells in blood and tissues. This, in turn, alleviates symptoms such as fever, nausea, fatigue, muscle weakness, and pain. Vaccines, including inactivated viral, live attenuated viral, protein subunit, viral vectored, mRNA, DNA, and nanoparticle vaccines, significantly reduce the adverse long-term effects of the virus. However, no vaccine has been reported to provide lifetime protection against COVID-19. Consequently, protective measures such as physical distancing, mask usage, and hand hygiene remain essential strategies. This review offers a comprehensive understanding of the persistent effects of COVID-19 on individuals of varying ages, along with insights into diagnosis, treatment, vaccination, and future preventative measures against the spread of SARS-CoV-2.

**Graphical abstract**

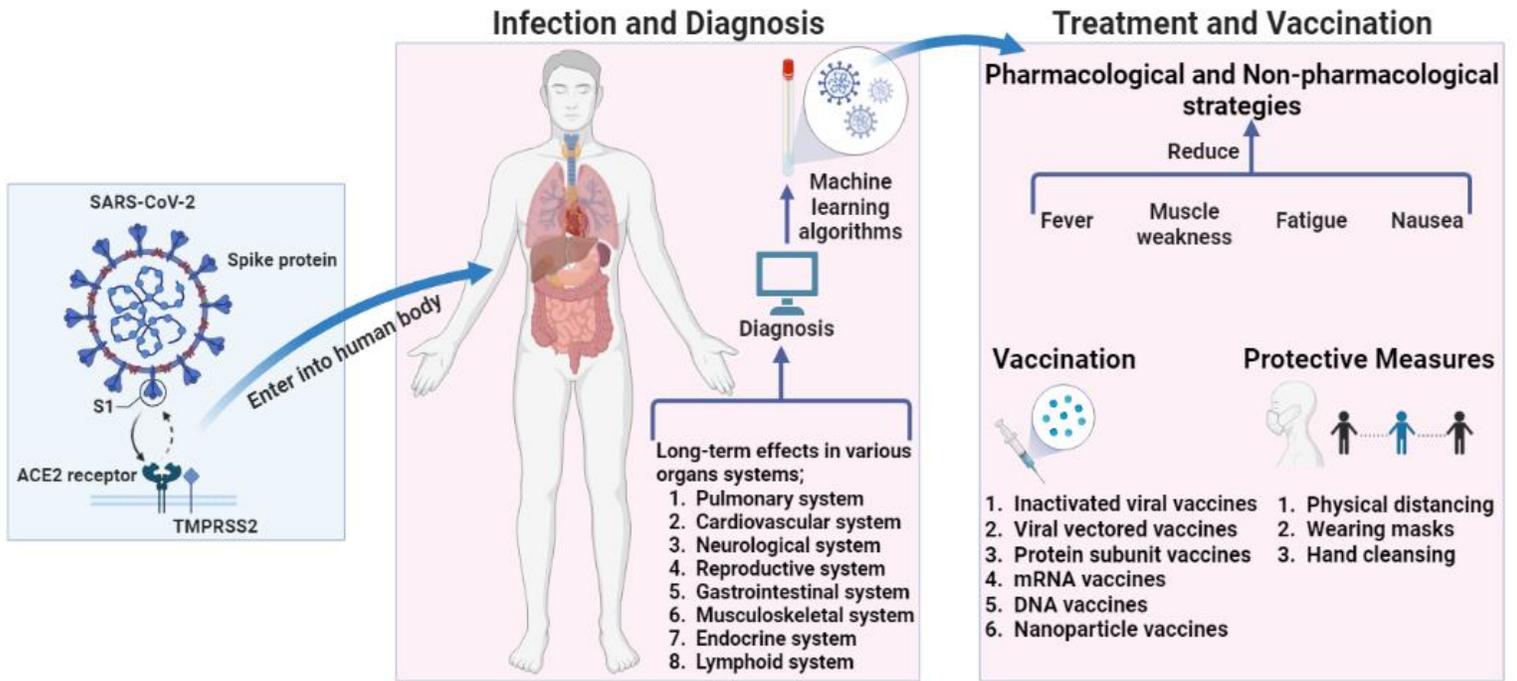

COVID-19: post infection implications in different age groups, mechanism, diagnosis, effective prevention, treatment, and recommendations (generated by using BioRender.com).

## 1. Introduction

The pneumonia outbreak in Wuhan Cityof at the beginning of December 2019 was caused by a novel pathogenic virus from the Coronaviridae family called coronavirus (CoV), also recognized as '2019 novel coronavirus (2019-nCOV) [1,2]. According to Kaniyala Melanthota et al. [3] "corona" is a Latin word that means "crown" or "halo" because of its characteristic appearance when viewed under a microscope. Coronaviruses are a group of viruses with a distinct structure consisting of a crown-shaped spike protein on their surface, which enables them to attach to and penetrate host cells (figure 1) [4–6].

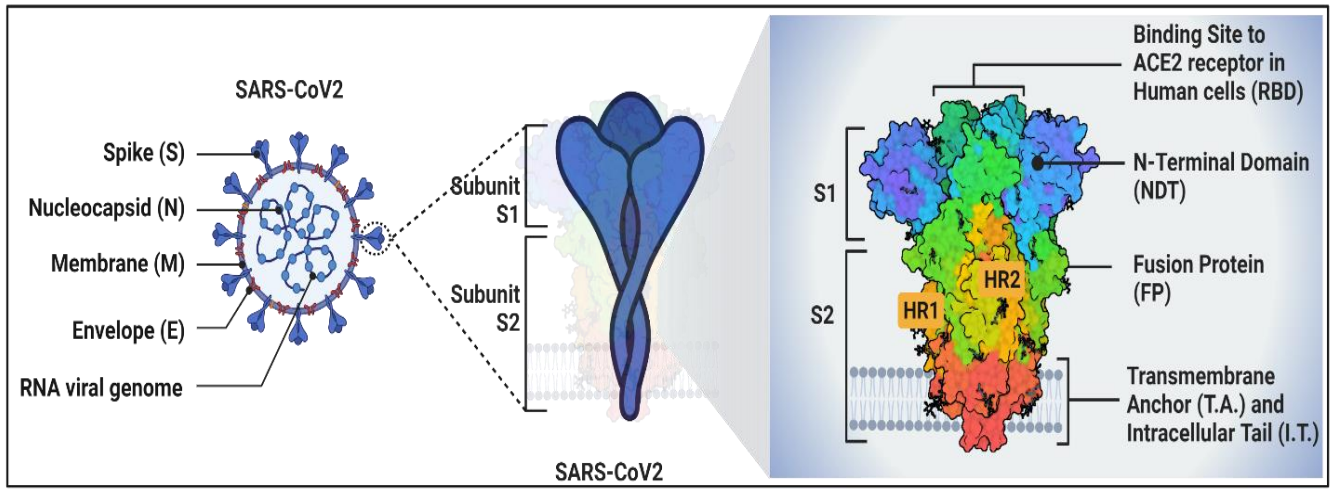

**Figure 1:** The SARS-CoV-2 spike protein is crucial to infects human cells. It consists of three main components: the S1 subunit; recognize and bind to the host cell receptor, which is the angiotensin-converting enzyme 2 (ACE2), the S2 subunit; facilitates the fusion of the viral and host cell membranes, and the receptor-binding domain (RBD); region within the S1 subunit that directly interacts with the ACE2 receptor on the surface of human cells (generated using BioRender.com).

Coronaviruses primarily induce infections of the respiratory and gastrointestinal tracts and are classified into 4 main kinds: Alphacoronavirus, Betacoronavirus, Deltacoronavirus, and Gammacoronavirus [7,8]. The first 2 kinds infect mammals predominately, while the last two infect birds chiefly [9]. Humans are the most susceptible mammalian species to coronavirus, with approximately 677 million cases and 6.9 million fatalities recorded by March 2023 [10]. There are approximately seven recognized types of human coronaviruses namely HCoV229E, HCoV-NL63, HCoVHKU1, HCoV-OC43, Middle East respiratory syndrome coronavirus (MERS-CoV), severe acute respiratory syndrome coronavirus (SARS-CoV) and severe acute respiratory syndrome coronavirus 2 (SARS-CoV-2) [11]. The first two kinds (HCoV229E and HCoV-NL63) belong to the Alphacoronavirus, while the remaining five (HCoVHKU1, HCoV-OC43, MERS-CoV, SARS-CoV and SARS-CoV-2) are Betacoronaviruses [1,12]. Coronaviruses garnered global attention in the 2003 pandemic of SARS, preceded by the 2012 epidemics of MERS and COVID-19 [13,14]. SARS-CoVand MERS-CoV are recognized as extremely pathogenic viruses that are transmitted from bats to dromedary camels and humans [15].

SARS-CoV-2 is a highly contagious and pathogenic coronavirus primarily responsible for the COVID-19 pandemic [16,17]. Dust particles and fomites transmit COVID-19 during close, unprotected contact between an infector and an infected individual [12]. Based on empirical evidence, airborne transmission of COVID-19 has not been documented and is not known to be a significant transmission mechanism, although it is conceivable that such aerosol-generating practices may be employed in medical facilities [18–20]. COVID-19 spreading through fecal is also reported in a few clinical investigations in such a way the active virus was observed microscopically in the patient's feces [21,22]. In addition, the fecal-oral route is also responsible for transmitting COVID-19; for example, when Hindson [23] performed a clinical characterization of ten pediatric patients with SARS-CoV-2 infection in China, none of whom required intensive care because they all lacked pneumonia symptoms, eight tested positives on rectal swabs despite negative nasopharyngeal testing. Therefore, the author concluded that the digestive system may be more susceptible to viral dissemination than the respiratory system.

After the successful transmission of SARS-CoV-2 to a suitable host, the spike protein (S protein), located on the virus's surface, assists the virus in infecting host cells [24]. The S protein comprises 1162 and 1452 amino acids [25,26]. S1 and S2 are two subunits of the spike protein with distinct functions. The global S1 subunit consists of two primary domains: the N-terminal domain (NTD) and the receptor-binding domain (RBD) or C-terminal domain (CTD), which permit the virus to recognize and attach to the angiotensin-converting enzyme 2 (ACE2) receptor on the surface of human cells [24,25]. The ACE2 gene encodes a protein of 805 amino acids and belongs to the angiotensin-converting enzyme family of dipeptidyl carboxydipeptidases. It is typically expressed in the cell membranes of cells in numerous human organs, including the heart, intestine, kidneys, lungs, testis, ovaries, and arteries [27]. The expression of the ACE2 receptor on various organs is crucial to the entry and infection processes of SARS-CoV-2 [28,29].

The S2 subunit is highly conserved and contains segments with crucial roles in virus-cell fusion. As shown in Figure 1, the S2 subunit contains the heptad repeat regions (HR1 and HR2), the transmembrane anchor (TA), the fusion peptide (FP), and the intracellular tail (IT) [24]. The FP performs an essential role in mediating the membrane fusion reaction by forming a brief helix in which the conserved hydrophobic elements are submerged at an interface with other S2 elements. TA is a hydrophobic region of the spike protein that adheres to the viral membrane. A polybasic amino acid bridge connects the S protein's two subunits. HR1 and HR2 are fusion protein regions that interact with one another during membrane fusion. HR1 forms a coiled-coil structure that interacts with the HR2 region to bring the viral and host cell membranes closer, facilitating fusion [30].

Pathogens' chronic or long-term effects are more detrimental than acute ones because acute diseases are short-lived and chronic diseases are long-lasting. The long-term effects of COVID-19 are defined as an inflammatory or host response to the virus that begins approximately four weeks after initial infection and persists for a duration that has not yet been determined [31,32]. In literature, many terminologies such as "long-haul COVID," "post-acute sequelae of SARS-CoV-2 infection (PASC)", "post-acute COVID-19 syndrome", "post-COVID syndrome," "chronic COVID," "long COVID," "long-term COVID," "long COVID-19", "persistent COVID-19," and "long-term sequelae of COVID-19" were generally referred to the ongoing or persistent symptoms that some individuals experience after their initial acute COVID-19 infection has resolved [33]. More than fifty-five general persistent symptoms in humans due to COVID-19 were reported by [34], including fatigue, headache, attention disorder, hair loss, and dyspnea. After 120 days of infection, these long-term effects of COVID-19 are most prevalent in Asians [35,36]. Globally, more than 200 million people suffer from the persistent effects of COVID-19 that disrupt their normal body functions [35,36]. In addition to the generalized persistent effects of COVID-19, more specific long-term effects on various organ systems and age groups must be investigated. Moreover, effective diagnosis, treatments, and vaccinations for COVID-19 are currently limited. Therefore, this review aims to report the long-term effects of COVID-19 on the cardiovascular, reproductive, neurological, gastrointestinal, pulmonary, musculoskeletal, hormonal, and lymphatic systems of humans of various age groups. Furthermore, current diagnosis, potential pharmacological and non-pharmacological treatment strategies, the effectiveness of different vaccines against COVID-19, and protective measures that limit the spreading of SARS-CoV-2 were also explored.

## 2. Review Methodology

Using subject-specific keywords (long-term COVID-19 effects, human systems, COVID-19 treatments, and COVID-19 vaccines), research articles and reviews were retrieved from ScienceDirect, Google Scholar, and PubMed. These databases contain many publications that have been peer-reviewed and are widely used. This technique selects publications published between 2018 and 2023.

## 3. Long-term effects of COVID-19 on the human systems

### 3.1. Pulmonary system

The pulmonary system is the initial interaction site of the SARS-CoV-2 with its host because a healthy individual inhales an infected person's exhaled droplet. The bronchial epithelium and lungs are the most vulnerable tissues because of the high numbers of ACE2 receptors on their cell membranes, which provide the site for virus infection. Damage to the alveoli of the human lung caused by SARS-CoV-2 also increased the permeability of the microvessels, which enhanced virus movement in the bloodstream and, ultimately, the heart via pulmonary circulation, resulting in cardiac tissue damage and thrombosis [37].

After inhalation of the SARS-CoV-2, the disruption of homeostasis begins in the lungs by primarily infecting alveolar type II and I pneumocytes that express ACE2, resulting in pneumocyte inflammation and injury. In response to a viral infection, immune cells such as resident macrophages and recruited monocytes release chemokines that attract other immune cells, such as mononuclear and polymorphonuclear cells, for phagocytosis. Moreover, adipocytes are fat cells that also behave as immune cells by producing proinflammatory substances such as tumor necrosis factor- (TNF-α) and interleukin-6 (IL-6) [38]. Monocytes and neutrophils are mononuclear and polymorphonuclear phagocytic cells, respectively, that secrete extracellular traps, which are net-like structures composed of DNA-histone complexes that bind pathogens [39,40]. In addition, the inflammatory process induces oxidative stress by generating free radicals that impair protein function [41]. The secretion of proinflammatory cytokines, extracellular traps, and oxidative stress thus causes thromboinflammation [42].

After 120 days of infection, the patients with severe pneumonia showed increased density of lung tissue and enhanced glycolytic activity due to an increased number of inflammatory cells [43], which may have accumulated through migration or local proliferation, along with cellular activation. Therefore, more inflammatory cells cause the elevation of adhesion molecules, such as IL-6, angiopoietin-1, vascular cell adhesion molecule-1 (VCAM-1), and intercellular adhesion molecule-1 (ICAM-1), in the bloodstream [44]. These molecules work as the activation of endothelial cell, but overproduction of these molecules cause endothelial cell dysfunction; ultimately, endothelial cells are more prone to SARS-CoV-2 infection and cause long COVID-19 [45]. Additionally, after 365 days of initial infection in adult, various abnormalities in the lung were observed, including elevations of TNF, C-reactive protein (CRP), interleukin-β (IL-β), and IL-6, indicating persistent systemic inflammation [46].

Endothelial cells contain Weibel-Palade bodies containing glycoproteins, p-selectin, and Von Willebrand Factor (VWF) [47]. VWF binds to collagen fiber beneath the endothelium and lines the interior of blood vessels, serves as a platform for the activation of factors involved in blood clotting and promotes the adhesion of substances such as fibrin, leukocytes, and platelets, and ultimately exacerbates inflammation and excessive blood clotting, resulting in hypercoagulability (figure 2) [43]. In addition, endothelial cells are responsible for releasing E-selectin from their exterior membrane, thereby adhering leukocytes to the blood vessel wall at the site of inflammation. Consequently, these results indicated that perturbation of homeostasis causes in inflammation and immune activation that persists for an extended period following the initial SARS-CoV-2 infection [48].

Long-term follow-ups of coronavirus survivors revealed persistent impairments in pulmonary function tests (PFT) years after recovery [49]. To investigate this statement, Salem et al. [50] recruited fifty subjects aged 18–60 years and divided them into two groups: the control group (30 subjects) and the post-COVID-19 pneumonia group (20 patients). Results indicated that 50 % of post-COVID-19 cases were found to have restrictive lung impairment, compared to 20 % of the control group. Also, the long-term effects of SARS-CoV-2 on lung function showed that the diffusing capacity for carbon monoxide ($DL_{co}$), forced vital capacity (FVC), and total lung capacity (TLC) were all significantly lower in post-COVID-19 subjects (3 months after recovery) than in controls. The harmful, persistent effects of SARS-CoV-2 in patients might be improved in case of six months of respiratory rehabilitation; for instance, Liu et al. [51] conducted a randomized controlled trial with 72 participants (older adults aged 65 years or above). Results indicated that Six weeks of respiratory rehabilitation, including daily 6-minute' walks, significantly improved respiratory function and the quality of life and reduced anxiety in elderly patients.

The severity of SARS-CoV-2 infection in the pulmonary system of humans is identical in both sexes (male and female) [52,53]. Thus, Percivale et al. [54] conducted a study with 117 females and 214 males of different age groups (0-39, 40-61- and $\geq 61$ years) affected by SARS-CoV-2 pneumonia. They found that the results of computed tomography (CT) of male and females' chest of the same age distribution, have no statistically significant differences been observed regarding CT scan findings, such as increased lung opacity and highly enlarged mediastinal lymph nodes. A previous study by Dangis et al. [55] also revealed that the same lung injury was observed after recovery from coronavirus in elderly males and female COVID-19 patients (65 years old). Various other long-term effects of SARS-CoV-2 on pulmonary system of various age groups are detailed in Table 1.

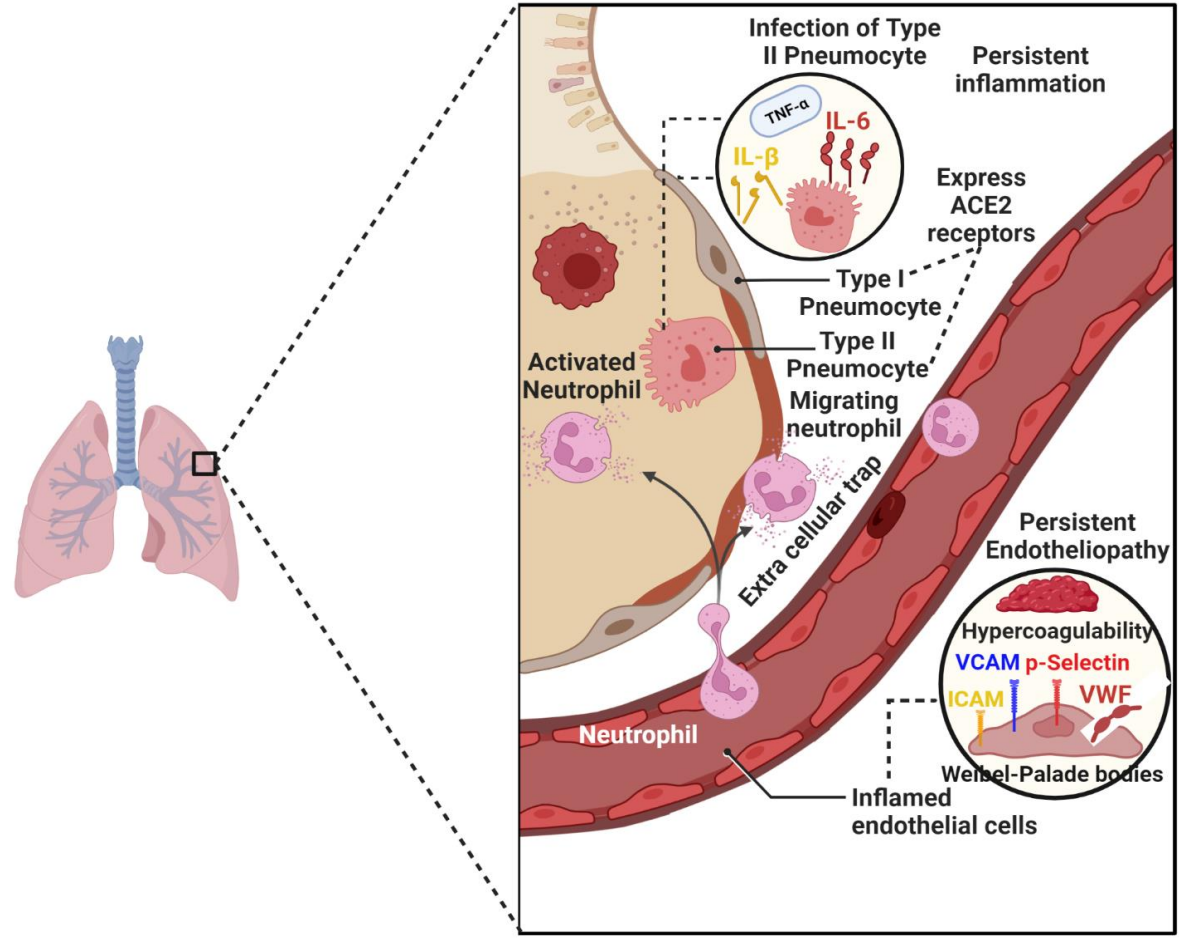

**Figure 2:** Persistent effect of COVID-19 on alveolus: The virus mainly targets and infects two types of cells in the alveoli called type I and type II pneumocytes. These cells are crucial for the proper functioning of the lungs. Virus triggers an ongoing inflammatory response in the alveoli, even after the acute phase of the infection has passed. This persistent inflammation can lead to tissue damage and impair lung function. Elevated levels of adhesion molecules in the bloodstream such as IL-6, angiopoietin-1, VCAM-1, and ICAM-1, are found at increased levels in the blood of individuals with COVID-19. These molecules serve as markers indicating activation of endothelial cells, which are cells lining the blood vessels (generated by using BioRender.com).

### 3.2. Cardiovascular system

Globally, 40.9 % of COVID-19 patients are diagnosed with myocardial injury; therefore, the long-term effect of COVID-19 on the cardiovascular system remains a worldwide concern [56,57]. Generally, cardiac cells expressing high ACE2 levels are more susceptible to SARS-CoV-2 invasion and subsequent organ injury, including cardiomyocytes, and the adult human heart has more ACE2 expression than the lung [58]. The transmembrane serine protease (TMPRSS2) is a protein responsible for the virus's cleavage into S1 and S2 subunits to facilitate the recognition and binding with ACE2 receptor. SARS-CoV-2 enters cardiomyocytes via an endosomal-dependent pathway that does not require TMRSS2 because numerous endosomal proteases, such as cathepsin and calpains, are responsible for virus entry (figure 3) [59]. Moreover, Omicron, a new SARS-CoV-2 variant, only requires ACE2 to invade the host cell without the help of TMPRSS2 [60].

Myocardial contractile dysfunction is among the most persistent effect of COVID-19. Because SARS-CoV-2 directly destroys sarcomeres, the fundamental elements of cardiomyocytes consist of two primary myofilaments responsible for muscular contraction [61]. The two primary myofilaments consist of thick filaments made of the protein myosin and thin filaments made primarily of the protein actin [62]. The myosin protein of thick filament has the LKGGK sequence, which is the preferred site for the SARS-CoV-2 virus to cleave the thick filament with its papain protease, ultimately resulting in sarcomere destruction [63]. In addition, the virus also decreased the expression of sarcomeric genes such as MYL2 and MYH6 that retard sarcomere function [64]. Furthermore, CDKI and IDH2 are genes involved in glycolysis and oxidative phosphorylation in mitochondria [65], two major metabolic pathways for energy production. SARS-CoV-19 is responsible for the reduction of expression of these genes (CDKI and IDH2) [66], resulting in a reduction in energy production that adversely affects the contraction of the cellular and tissue levels, resulting in oxidative stress and inflammation of cardiomyocytes [67]. The SARC-CoV-2 virus releases non-structural protein 1 (Nsp1), which binds to the 40S ribosomal subunit in such a way that the C-terminal domain directly blocks messenger RNA entry in the ribosome, perturbing the host cell by shutting down the translation of host messenger RNA, and consequently impairing myocardial contractility (figure 3(1))[67]. In addition to disrupting the mechanical function of cardiomyocytes, SARC-CoV-2 also disrupts the electrical function by inducing structural changes in cardiomyocytes by forming syncytia [68]. The sarcolemmal t-tubule viral S protein forms intracellular junctions between cardiomyocytes as a result of the cell-to-cell fusion of non-infected cells caused by the fusogenic property of the SARC-CoV-2 S protein [69,70]. The formation of junctions promotes electrical short circuits in the myocardium and results in electrophysiological abnormalities, such as an imbalance in $Ca^{2+}$ flux and sparks (figure 3(2)) [71].

The SARS-CoV-2 infection is also responsible for the loss of ACE2 in cardiomyocytes. ACE2 is a homology of angiotensin-converting enzyme (ACE) and negatively regulates the renin-angiotensin-aldosterone system (RAAS) by breaking down a molecule, Angiotensin II (AngII), into angiotensin (1–7) (Ang (1–7)), while ACE converts AngI into AngII [71]. The S protein of the SARS-CoV-2 virus, once it binds to ACE2, induces ACE2 shedding with the help of a transmembrane protease, a disintegrin, and metalloproteinase-17 (ADAM-17) [72], resulting in ACE2 being unable to break AngII down into Ang (1–7). As a result, AngII accumulates, which promotes harmful processes like excessive autophagy (cellular self-degradation) and apoptosis (cell death) in cardiomyocytes. These processes are believed to contribute to the development of heart complications such as thrombosis and hypertension, which are more common in older adults (65 years or older) [72]. The results showed that SARS-CoV-2's persistent mechanical and electrical disruption of cardiomyocytes causes thrombosis and hypertension (Figure 3(3)).

Hypertension and coronary heart disease are the most common complications based on the long-term effects of COVID-19 [73,74]. According to the Italian Public Institute, The Istituto Superiore di Sanita, after the COVID-19 outbreak, about 68.3 % of patients were diagnosed with hypertension, followed by 28.2 % with ischemic heart disease and 22.5 % with atrial fibrillation [75]. Hypertension is associated with severe cardiomyocyte functional loss because cardiomyocytes depend on the endothelial cells not only for oxygenated blood supply [76], and endothelial cell dysfunction is majorly due to aging and decline of sex hormones during aging [77]. Bucciarelli et al. [78] claimed that gender is a potential modifying factor in cardiovascular disease cases involving long-term COVID-19 complications. Furthermore, recovered females from COVID-19 experienced more long-term complications regarding the cardiovascular system than males [78], such as arrhythmia, acute coronary syndrome, heart failure, hypertension, myocardial fibrosis, right ventricular dysfunction and diabetes mellitus [79]. Moreover, Dhaibar and Cruz-Topete [80] revealed the sex-specific effects of cortisol on cardiovascular health in young and middle-aged females recovering from COVID-19. Chronic cortisol production, triggered by fear of reinfection, significantly compromised women's cardiovascular health. Additional complications across different age groups are outlined in Table

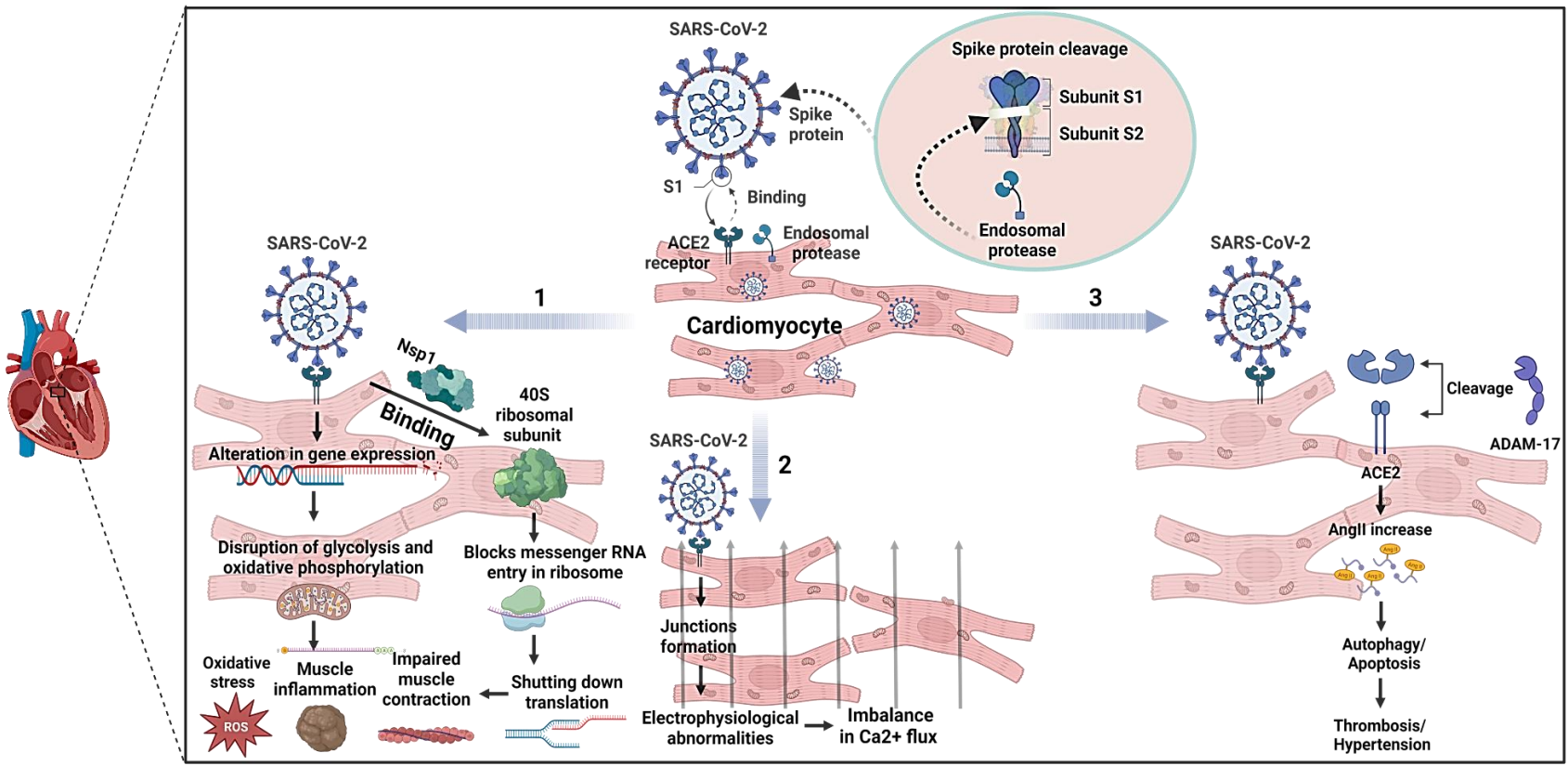

**Figure 3:** The persistent effect of COVID-19 on cardiomyocytes; SARS-CoV-2 enters cardiomyocytes by exploiting the endosomal pathway. The virus binds to the cell membrane of cardiomyocytes and being engulfed into the cell within a vesicle called an endosome. The endosomal protease enzyme causes the cleaves of the spike protein (S protein) of the virus. This cleavage is crucial for the virus to release its genetic material into the host cell. The cleavage, exposed S1 subunit of the virus specifically recognizes and binds to the angiotensin-converting enzyme 2 (ACE2) receptor, and causes direct cardiomyocyte injury through three mechanisms. (1) When Nsp1 binds to the ribosomal 40S subunit, it triggers the cessation of translation, consequently impairing cardiomyocyte muscle contraction. Additionally, disruption of glycolysis and oxidative phosphorylation leads to muscle inflammation and oxidative stress. (2) The formation of syncytia generates junctions that induce electrophysiological abnormalities. (3) Cleavage of ACE2 by ADAM-17 results in elevated levels of AngII, which in turn promotes detrimental processes like cellular self-degradation and cardiomyocyte death (generated by using BioRender.com).

### 3.3. Neurological system

After the COVID-19 pandemic, persistent neurological effects have become a significant cause of morbidity and mortality [81]. The olfactory nerve is a route for SARS-CoV-2 to enter the brain [82,83]. This way, the olfactory bulb and sustentacular cells in the olfactory epithelium highly express ACE2 receptors and TMPRSS2 proteins and provide a binding site for the virus [84]. The olfactory bulb structure is found in the cerebral hemisphere, a part of the brain where viruses are transported to other parts by anterograde and retrograde pathways through cerebral neurons [85]. The neuronal membrane (the outer membrane of the neuron) of cerebral neurons is also enriched with ACE2 receptors, where the S protein of SARS-CoV-2 binds. Through endocytosis, the virus enters the neuron, combines with the microtubule, and transports anterogradely from the cell body to the axon [85]. Moreover, the virus transfers trans-synaptically through retrograde transport from the axon of one neuron to the cell body of another neuron and causes chronic neuropathology in the central nervous system [82–84]. Rapid retrograde or anterograde transport of SARS-CoV-2 through microtubules invades the central nervous system and disrupts brain signaling (Figure 4) [86]. Moreover, the blood-brain barrier is another route by which SARS-CoV-2 enters the brain via basement membrane disruption (Figure 4) [87]. The blood-brain barrier mainly consists of endothelial cells connected by the tight junction, and a basement membrane embeds the brain endothelial cells and astrocytes due to the high expression of ACE2 receptors in endothelial cells allowing SARS-CoV-2 to penetrate the blood-brain barrier. Furthermore, the matrix metalloprotease-9 (MMP-9) protein in the basement membrane restricts the activity of pathogens [87]. However, due to the high expression of ACE2 receptors on the basement membrane, the virus S protein attaches to the membrane, alters the expression of MMP-2 protein, makes the membrane permeable for virus entrance, and then passes to the astrocytes [28]. Many neurons surround the astrocytes, and the outer membrane of a neuron contains ACE2 receptors; therefore, virus entry into neurons is unrestricted, ultimately invading the central nervous system and disrupting brain signaling [88].

After SARS-CoV-2 infected the human nervous system, the first long-persistent symptoms after the patient's recovery were smell and taste perception alterations. According to the study conducted by Yan et al. [89], 20.21 % of infants with COVID-19 exhibited taste and olfactory dysfunctions, and adolescents in Asia are highly resistant to this long-term effect of COVID-19. Other common long-term effects include headache, nausea, impairment of short-term memory, and inattention, which are most common in elderly COVID-19 survivors [90–94]. In addition to these general persistent neurological effects of COVID-19 on humans, some life-threatening consequences of SARS-CoV-2 invasion into the central nervous system include encephalitis, brain inflammation, seizures, fever, and fluctuations in electroencephalography patterns [95,96]. Furthermore, as a result of blood-brain barrier dysfunction, persistent stroke in elderly COVID-19 survivors was observed to be prevalent and positively correlated with advancing age [97]. Siow et al. [98] conducted a systematic analysis and found that over 70.5 % of COVID-19 patients with acute infection suffered a stroke as a long-term complication. Other complications regarding human neurological system   of various age groups are described in Table 1.

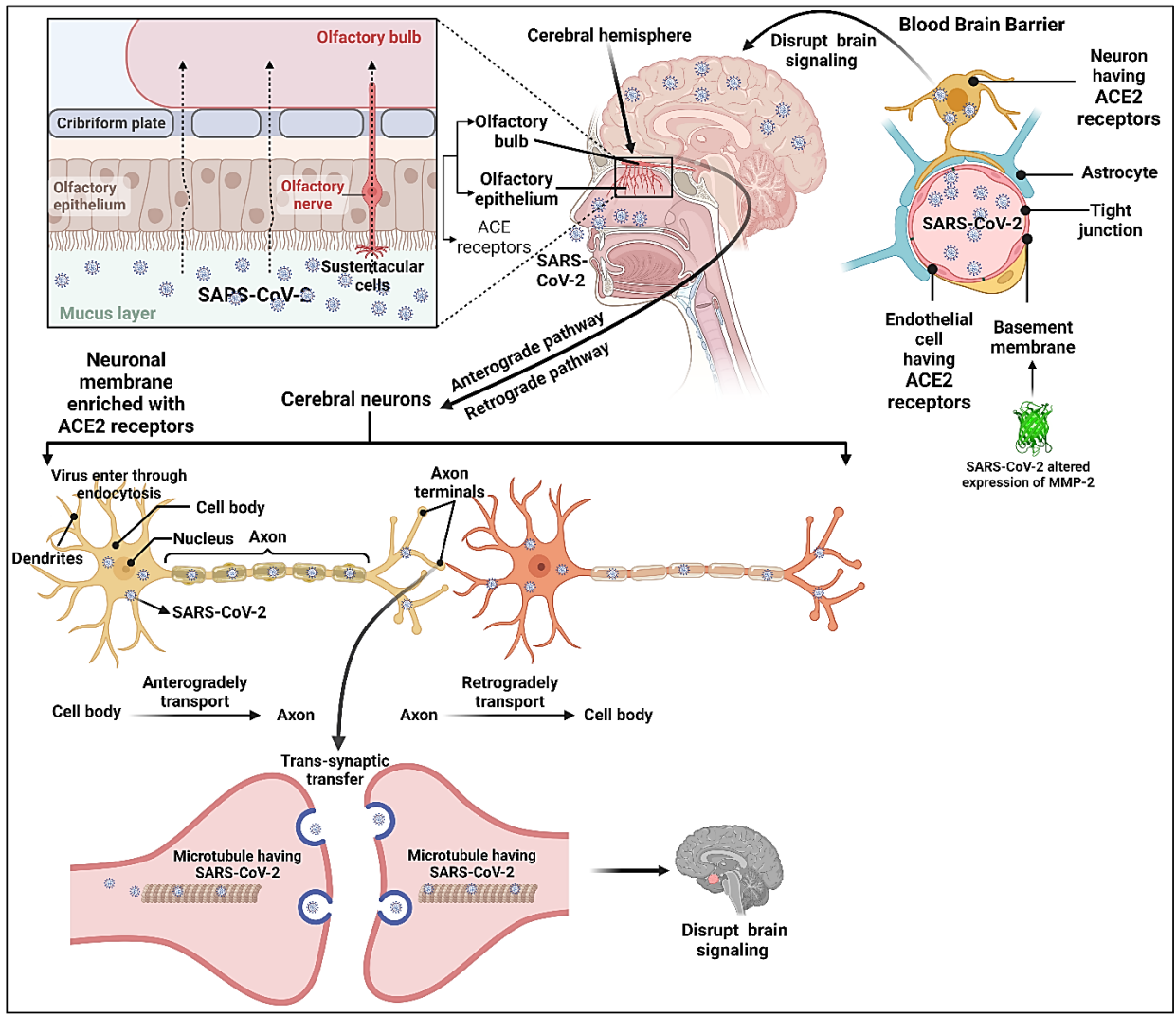

**Figure 4:** SARS-CoV-2 can infiltrate the central nervous system (CNS) through two primary pathways: the olfactory bulbs and the blood-brain barrier. In the olfactory pathway, the virus enters via inhalation, infects the olfactory epithelium in the nasal cavity, and travels along the olfactory nerve to the olfactory bulb in the brain. This invasion disrupts normal brain signaling, potentially causing neurological symptoms. Alternatively, SARS-CoV-2 can breach the blood-brain barrier, usually protecting the CNS from bloodstream contaminants, by disrupting the basement membrane of brain capillary endothelial cells. Once the virus gains access to the brain parenchyma, it can infect brain cells directly, leading to neurological complications (generated by using BioRender.com).

### 3.4. Reproductive system

The male and female reproductive systems play crucial roles in the human body and are essential for the species' survival [99]. Males are more prone to SARS-CoV-2 infection than females, and the fatality rate is also higher in males because the testes have more ACE2 receptors than the ovaries [100]. Four various testicular cells, such as the seminiferous tubule, spermatogonial stem cells, Leydig cells, and myoid cells, are showing high expression of ACE2. The expression of ACE2 protein is age-specific in males, with a high positive rate at the age of 30, indicating that SARS-CoV-2 can infect young males more than older ones. Furthermore, infertile males are more susceptible to SARS-CoV-2 infection than fertile normal males because they have more ACE2 receptors [101]. The co-expression of the ACE2 receptor and the entry-associated protease TMPRSS2 in spermatogonial stem cells, Leydig cells, and myoid cells is very high and provides a binding site to the virus [102]. Due to virus invasion on Leydig cells (figure 5), testosterone production was reduced, which significantly caused chronic effects like spermatogenic dysfunction, such as reduced sperm count [103]. Moreover, male reproductive dysfunction due to COVID-19 poses persistent seminiferous tubule injuries, such as inflammation of the testis, known as orchitis. The persistent effects of COVID-19 on the male reproductive system are highly aging-dependent; as the individual ages, the Leydig cells and seminiferous tubule become weak, and the virus easily invades the cells, significantly affecting testis health [104]. Other persistent effects of SARS-CoV-2 on male reproductive system are described in Table 1.

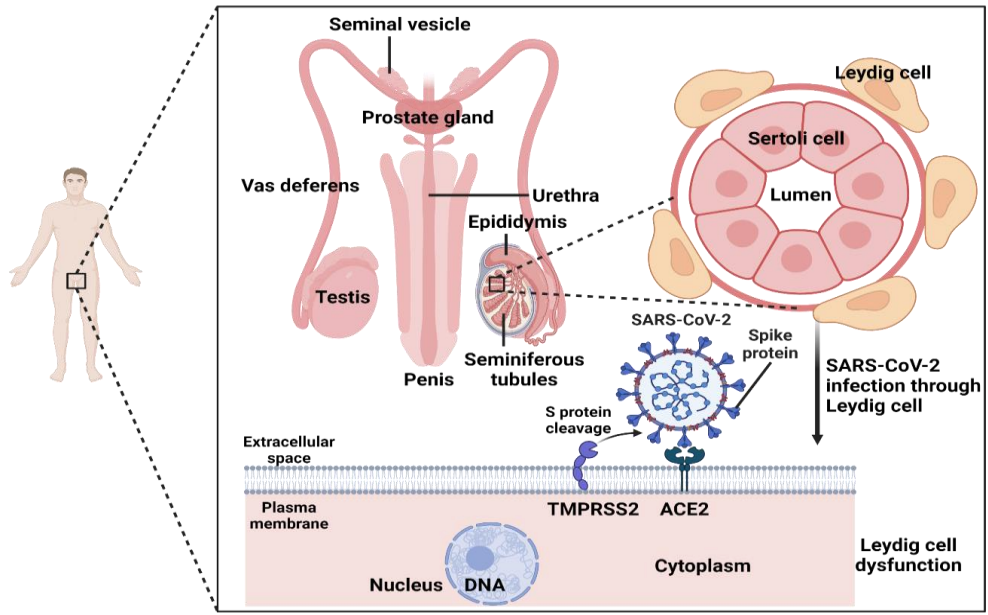

**Figure 5:** SARS-CoV-2 infects cells via the ACE2 receptor, abundant in the testes, facilitated by the enzyme TMPRSS. Both are highly expressed in Leydig cells, responsible for testosterone production. Co-expression in Leydig cells makes them susceptible to SARS-CoV-2 infection, potentially leading to impaired sperm production and hormonal imbalance, affecting male reproductive health (generated by using BioRender.com).

In addition to being expressed in male reproductive organs, ACE2 receptors are also present in female reproductive organs such as the ovaries and uterus. Oocytes (eggs) are produced by the ovaries for fertilization and are contained in follicles, and after oocytes maturation, the follicles release them [105]. Primarily, follicles are densely packed shells of somatic cells designed to protect the oocytes during development. During oocyte development, various types of follicles, such as primordial (primary or secondary) follicles and antral (tertiary) follicles, protect oocytes. The co-expression of ACE2 receptors and the entry-associated protease TMPRSS2 increases with oocyte maturation: 17 % in the primordial follicle, 39 % in the secondary follicle, and 62.1 % in the antral follicle [99]. After cleavage of the S protein of SARS-CoV-2 by TMPRSS2 and binding the S protein with ACE2, the virus enters antral follicles and inhibits follicle development; resultantly, oocyte quality is compromised (figure 6) [106]. The oocytes are a key determinant of ovarian reserve, and diminished ovarian reserve may adversely affect female fertility [107]. Therefore, SARS-CoV-2 infection diminishes ovarian reserve and oocyte quality, resulting in infertility and miscarriage in females [108]. Furthermore, the number of growing follicles decreases with aging, and SARS-CoV-2 infection exaggerates the decreasing antral follicle development that adversely affects the female reproductive system [109]. Li et al. [107] stated that at the age of 35 years, normal female fertility is reduced due to diminished ovarian reserves and oocyte quality. In contrast, for females who recovered from COVID-19, their fertility is reduced before 35 years of age [107]. Soysal and Yilmaz [110] performed a prospective cross-sectional study on sixty adult female patients. They discovered that the female recovered after COVID-19 caused persistently low levels of anti-mullerian hormone (AMH). AMH is a hormone corresponding to an individual's egg count and a parameter to analyze ovarian reserve. Besides the disturbance of the ovulation process due to SARS-CoV-2, the virus also affects the proliferative phase of the menstrual cycle [111]. In this proliferative phase, an endometrial layer of the uterus is highly expressed with ACE2 receptors and TMPRSS2 protein [112]. Therefore, the ACE2 and TMPRSS2 are positively correlated in the proliferative phase than in the early phase of the menstrual cycle [113]. In this context, Henarejos-Castillo et al. [111], who examined the gene expression in 112 patients with normal endometrium collected throughout the menstrual cycle, found that the expression of TMPRSS2 and ACE2 increases with age during the proliferative phase compared to other phases (ovulation and luteal phases) and hence provides more binding sites to the SARS-CoV-2 virus for invading the endometrium and ultimately making the female infertile. Table 1 describes additional long-term effects of SARS-CoV-2 on the female reproductive system.

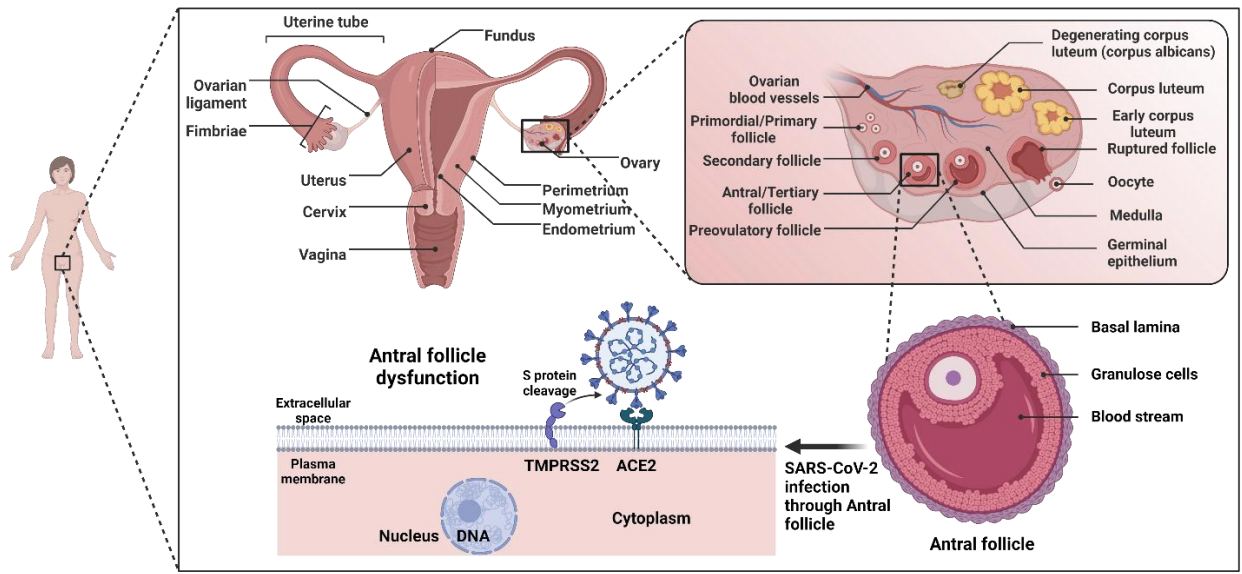

**Figure 6:** ACE2 receptors and TMPRSS2 are proteins found in cells. The SARS-CoV-2 virus uses ACE2 to enter cells, and TMPRSS2 is an enzyme that assists by priming the viral spike protein. Both ACE2 receptors and TMPRSS2 are present in antral follicles, providing potential binding sites for the SARS-CoV-2 virus. Antral follicles are structures in the ovaries that contain developing eggs and are crucial for the ovulation process. The presence of SARS-CoV-2 in antral follicles could lead to chronic ovulatory

dysfunction, which refers to persistent issues with the normal release of eggs from the ovaries during the menstrual cycle (generated by using BioRender.com).

### 3.5. Gastrointestinal system

SARS-CoV-2 enters the gastrointestinal tract via saliva or contaminated food [114] and causes gastrointestinal injury via direct cytotoxic damage, dysregulation of RAAS, or malabsorption of tryptophan in the intestinal epithelium [115]. There are one hundred times more ACE2 receptors in the gastrointestinal tract, particularly in the colon than in the alveolar cells of the lung and respiratory system, which serve as virus-binding sites [116,117]. The cleavage of the S protein of SARS-CoV-2 by TMPRSS2 renders the gastrointestinal epithelium susceptible to viral infection [118]. Pyroptosis and apoptosis directly damaged cells after SARS-CoV-2 infected the epithelium, leading to the epithelium's injury and inflammation (figure 7) [119,120]. In addition to its function as a receptor for viruses, ACE2 is a negative regulatory enzyme in the RAAS [121]. In addition, in RAAS, ACE cleaves Ang I to Ang II, which binds to the Ang type 1 receptor (AT1R) and mediates systemic processes, including inflammation and fibrosis [122]. On the other hand, ACE2 catalyzes the conversion of Ang II to Ang (1–7), and then Ang (1–7) binds to the G protein-coupled receptor Mas to mediate anti-inflammatory and anti-fibrotic effects [123]. The Ang II/AT1R signaling also promotes an immune response, ACE2 regulates immune functions via the Ang (1–7)/Mas signaling, and the balance between the Ang II/AT1R signaling and the Ang (1–7)/Mas signaling determines the overall homeostatic effect of the RAAS [124]. The S protein of SARS-CoV-2 binds to ACE2, and the downregulation of ACE2 promotes RAAS imbalance by increasing Ang II/AT1R signaling, decreasing Ang (1–7)/Mas signaling, and increasing Ang II level, ultimately leading to epithelium colitis injury (figure 7) [125]. Furthermore, besides causing cytotoxic injury and dysregulation of RAAS, SARS-CoV-2 also interferes with the absorption of essential amino acids in the human intestine [126]. ACE2 also regulates amino acid transport in the intestine and contributes to tryptophan absorption by binding to the membrane-bound amino acid transporter [53], which is called neutral amino acid transporter B $^{(0)}$ that are divided into two types B $^{(0)}$AT1 (SLC6A19) and B (0)AT2 (SLC6A15). The amino acid tryptophan is essential for intestinal homeostasis [127]. SARS-CoV-2 S protein binds to the extracellular domain of ACE2. It can result in the internalization of ACE2/B$^0$AT1 which reduces the uptake of tryptophan and deactivates the mammalian target of rapamycin (mTOR) [128,129]. mTOR is a pathway that controls the expression of antimicrobial peptides that influence intestinal microbiota composition.

Nevertheless, due to the SARS-CoV-2, its deactivation reduces the production of antimicrobial peptides, modifies the intestinal microbiota, makes it more susceptible to pathogens, and causes inflammation (figure 7) [115].

Prolonged intestinal infection, persistent endothelial injury, and microthrombi are the long-term effects of SARS-CoV-2 in patients after discharge [130]. Cheung et al. [131] detected the SARS-CoV-2 nucleocapsid protein in the appendix, colon, and ileum of five patients after 180 days of recovery from COVID-19. The enterocytes are columnar epithelial cells that line the inner surface of the small and large intestines and are permeable for the uptake of ions, water, nutrients, and vitamins [132]. SARS-CoV-2 infection causes disruption of intestinal permeability by altering the expression of the retinol-binding protein 2 (RBP2) gene [133] that suspends the absorption of retinoid compound (a derivative of Vitamin A) in the intestine [134]. Dibner [134] also claimed infection of enterocytes due to SARS-CoV-2 infection in patients after recovery from COVID-19. Aging is positively correlated with enterocyte dysfunction without discriminating the sex of patients [136,137]. Furthermore, hypochlorhydria, a condition in which gastric glands in the stomach do not produce gastric juices that, ultimately cause improper digestion due to SARS-CoV-2 infection and quite common in adult patients recover from COVID-19 [135,138].

The long-term symptoms of COVID-19 on the gastrointestinal system of patients after recovery include anorexia, ageusia, loss of appetite, abdominal distension, abdominal pain, diarrhea, vomiting and nausea [139,140]. The frequency of these long-term gastrointestinal symptoms in COVID-19 survivors ranged from 2 % to 79 %, and diarrhea most commonly persists in patients [141,142]. Moreover, Bogariu and Dumitrascu [143] verified the various persistent gastrointestinal symptoms due to COVID-19 syndrome in patients after twelve weeks of their initial infection based on the medical literature in the PubMed database. Results indicated that abdominal pain, diarrhea, nausea and vomiting are the main symptoms that exist in survivors. Previously, one more observational study conducted by Weng et al. [144] analyzed the long-term effects of COVID-19 in survivors after infection at twelve weeks and revealed that 24 % of patients showed loss of appetite followed by nausea (18 %), diarrhea (15 %), abdominal distension (14 %), belching (10 %), vomiting (9 %), abdominal pain (7 %) and rectal bleeding (2 %). Furthermore, 44.2 % of patients discharged from the hospital were sixty years old, 48.1 % were women, and 51.9 % were men who persistently acquired gastrointestinal symptoms due to COVID-19 [144]. More complications of gastrointestinal system due to SARS-CoV-19 is discussed in Table 1.

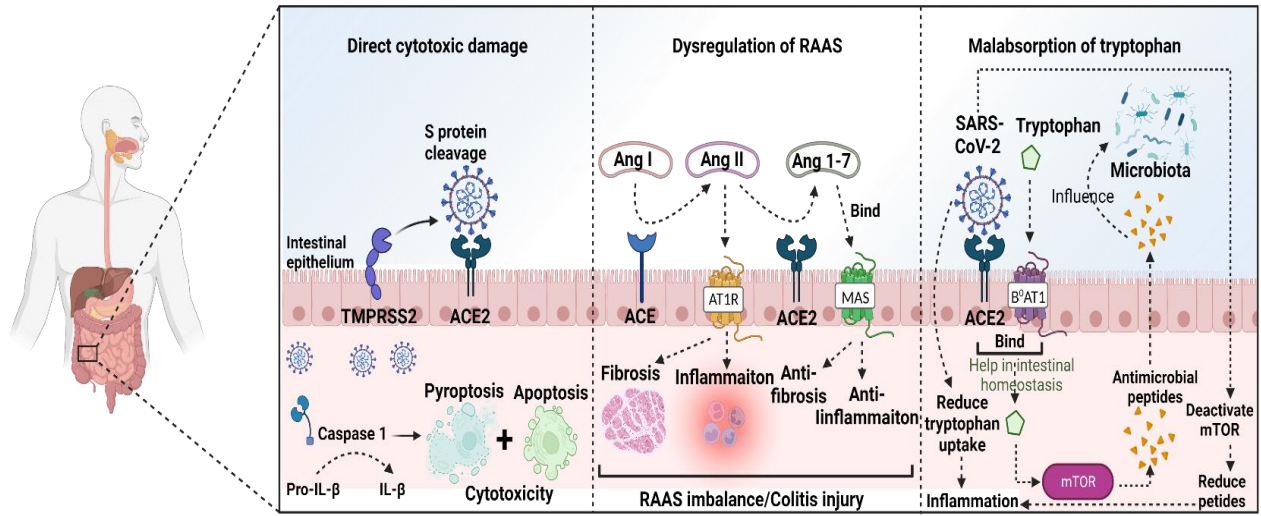

**Figure 7:** The gastrointestinal (GI) injury caused by SARS-CoV-2 can occur through three main pathways: Direct Cytotoxic Damage: SARS-CoV-2 can directly damage GI epithelial cells through pyroptosis and apoptosis, inflammatory forms of cell death triggered by infection, leading to tissue damage and cytokine release. Dysregulation of RAAS (Renin-Angiotensin-Aldosterone System): The virus's spike protein binds to ACE2 in GI cells, disrupting the RAAS balance. This imbalance, characterized by increased angiotensin II (Ang II) signaling through the angiotensin II type 1 receptor (AT1R) signaling and decreased Ang (1–7)/Mas signaling, contributes to colitis injury and inflammation. Malabsorption of Tryptophan: Interaction between the virus and ACE2 leads to internalization of ACE2 along with the amino acid transporter B0AT1, disrupting tryptophan absorption in the intestine. This affects immune regulation and antimicrobial peptide synthesis, exacerbating inflammation and GI injury, compounded by mammalian target of rapamycin (mTOR) pathway deactivation (generated by using BioRender.com).

### 3.6. Musculoskeletal system

40% of a human's body weight consists of skeletal muscle, composed of various myofibers and plays a mechanical function by converting chemical energy into mechanical energy, thereby generating movement [145,146]. In addition, skeletal muscle serves as a reservoir for the carbohydrates and proteins required for basal metabolism [147]. The dysfunction of musculoskeletal system due to SARS-CoV-2 through direct and indirect infection mechanism. In the direct infection mechanism, after cleavage of the S protein of the virus through TMPRSS2, it binds to the skeletal muscle cell surface that is enriched with ACE2 receptors and encounters the cell [118]. On the other hand, the SARS-CoV-2 infection is indirectly responsible for the inflammation of musculoskeletal muscles through the production of a "cytokine storm" by the human immune system (figure 8) [148]. Myokines are cytokines such as TNF-α, IL-6, IL-β, and IL-8 that are secreted by muscle fiber and ultimately disrupt tissue homeostasis by interfering with protein synthesis, resulting in metabolic dysregulation and retarding muscle growth (figure 8) [149,150]. The persistent atrophy of skeletal muscle is observed in patients recovered from COVID-19 after six months of being discharged from the hospital [151] by altering the expression of the peroxisome proliferator-activated receptor-gamma coactivator (PGC-1α) gene that also restrict the muscle contraction [152]. Furthermore, aging positively correlates to musculoskeletal atrophy [153]. Moreover, SARS-CoV-2 disrupts the regeneration of myofibers due to viral infection by altering the expression of the inositol-requiring enzyme1 α (IRE1α) gene [154] in elderly COVID-19 survivors [155].

The long-term effects of SARS-CoV-2 infection in patients revealed that elderly patients experienced a significant functional decline, such as diminished musculoskeletal fitness, promptly after discharge and at a twenty-four-month follow-up. Moreover, Xiong et al. [156] conducted longitudinal research on COVID-19 survivors in Wuhan, revealing that 28.3 % and 4.5 % of patients suffered from physical decline and myalgias, respectively. Due to the deposition of pro-inflammatory myokines in musculoskeletal tissues, elderly patients exhibit protracted general weakness and muscle fatigue due to muscle thickness for several weeks [157]. Even after protracted hospitalization, COVID-19 survivors typically retain functional disability, such as a diminished ability to walk and impaired balance, which ultimately impairs the performance of activities of daily living [158][151]. Greve et al. [159] also asserted that SARS-CoV-2 patients have long term effects after recovery due to musculoskeletal inflammation and muscle immobility. In general, sarcopenia positively correlates with aging, and elderly patients exhibited more musculoskeletal discomfort due to weakening of muscles [160]. Using previously published literature from PubMed, Embase, Cochrane Library, Scopus, and Web of Science, Xu et al. [161] systematically evaluated the prevalence of sarcopenia in COVID-19 survivors. The prevalence of sarcopenia was 48 % among 5,407 patients from 21 studies, with 42.5 % of sarcopenic patients being male and 35.7 % female. Some other musculoskeletal complications due to SARS-CoV-2 are detailed in Table 1.

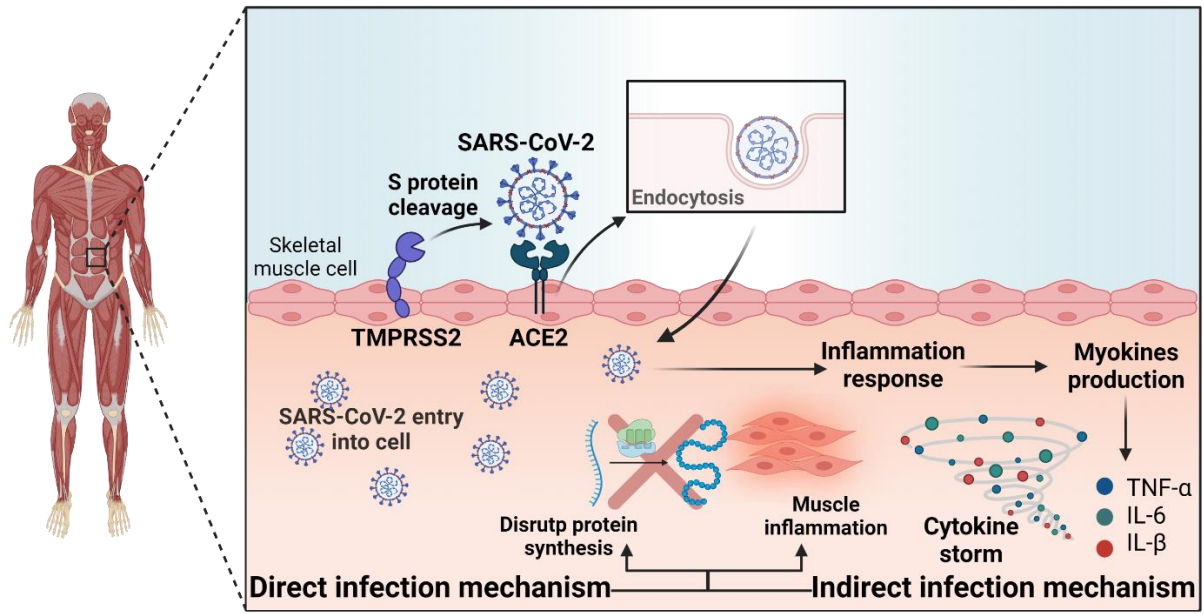

**Figure 8:** The musculoskeletal system can be affected by SARS-CoV-2 through both direct and indirect infection mechanisms: Direct Infection Mechanism: Skeletal muscle cells contain ACE2 receptors where the virus spike protein binds. This allows the virus to enter the cell and initiate infection. Once inside the cell, the virus can replicate, leading to damage and dysfunction of skeletal muscle tissue. Indirect Infection Mechanism: In response to SARS-CoV-2 infection, the immune system may produce excessive amounts of cytokines, resulting in a cytokine storm. This exaggerated immune response can lead to inflammation not only in the respiratory system but also in other tissues, including skeletal muscles. Additionally, the immune system may release myokines, which are cytokines produced by muscle cells. The production of myokines by the immune system can further exacerbate inflammation in muscle tissue, contributing to musculoskeletal dysfunction (generated by using BioRender.com).

### 3.7. Endocrine system

The endocrine system consists of glands that produce hormones that regulate development, growth, metabolism, tissue function, reproduction, etc. [162]. The endocrine system is vulnerable to SARS-CoV-2, and the olfactory nerve is a route for virus to enter the hypothalamus part of the central nervous system [83], as described in the earlier section (3.3). The pituitary gland, a "master gland" less than a pea size, is attached to the hypothalamus and is an integral part of the endocrine system because it secretes hormones (adrenocorticotropic hormone and thyroid-stimulating hormone) that regulate other important endocrine glands such as the adrenal and thyroid [163]. An adrenocorticotropic hormone is secreted by the pituitary gland, which stimulates the adrenal glands to secrete steroid hormones such as cortisol to regulate the body's stress response. Moreover, the pituitary gland has also released a thyroid-stimulating hormone into the bloodstream that stimulates the production of thyroid hormones such as thyroxine (T4) and triiodothyronine (T3) that maintain body homeostasis [164,165]. The pituitary glands are putative targets for the SARS-CoV-2 due to the expression of ACE2 receptors on the surface of their cells that ultimately disrupt the function of corticotroph cells and thyrotropic cells, resultantly reduced the production of adrenocorticotropic hormone and thyroid-stimulating hormone, respectively, therefore, adversely affect thyroid and adrenal glands (Figure 9) [166]. Gu et al. [167] performed a retrospective study involving forty-three patients hospitalized at Renmin Hospital of Wuhan University from February 5 to April 5, 2020, to investigate the potential long-term risks of SARS-CoV-2 on pituitary glands. The results indicated that adrenocorticotropic hormone production is reduced in COVID-19 survivors due to the dysfunction of corticotroph cells [167]. Furthermore, the virus also suppresses the function of thyrotropic cells, resulting in a reduction in the production of a thyroid-stimulating hormone that ultimately decreases thyroxine (T4) and triiodothyronine (T3) [167]. Wang et al. [168s] also performed a retrospective analysis on 84 hospitalized patients in the first affiliated hospital of Zhejiang University School of Medicine, Hangzhou, China, including eight hundred and seven healthy subjects as control. Results indicated that thyroid dysfunction was observed, including reduced thyroid-stimulating hormone, thyroxine, and triiodothyronine in a group of COVID-19 patients than the control group, and no significant difference was found in the age and sex of these two groups of subjects [168]. Long-term effects of COVID-19 on thyroid functions were also observed by [169], who indicated that thyroiditis was observed in all survivors without discriminating age and sex. Moreover, a case was reported about a COVID-19 survivor, 64 years old woman with hypothyroidism with a one-week history of abdominal pain, nausea, and vomiting. The survivor's diagnosis indicated the autoimmune disorder, Addison disease, in which more than 90 % of the adrenal cortex is destroyed, which cannot produce steroid hormones such as cortisol and causing fatigue [170]. Fatigue is the most prevalent symptom observed in patients who survive the SARS-CoV-2 infection on the thyroid and adrenal [171]. Other long-term symptoms in COVID-19 survivors are described in Table 1.

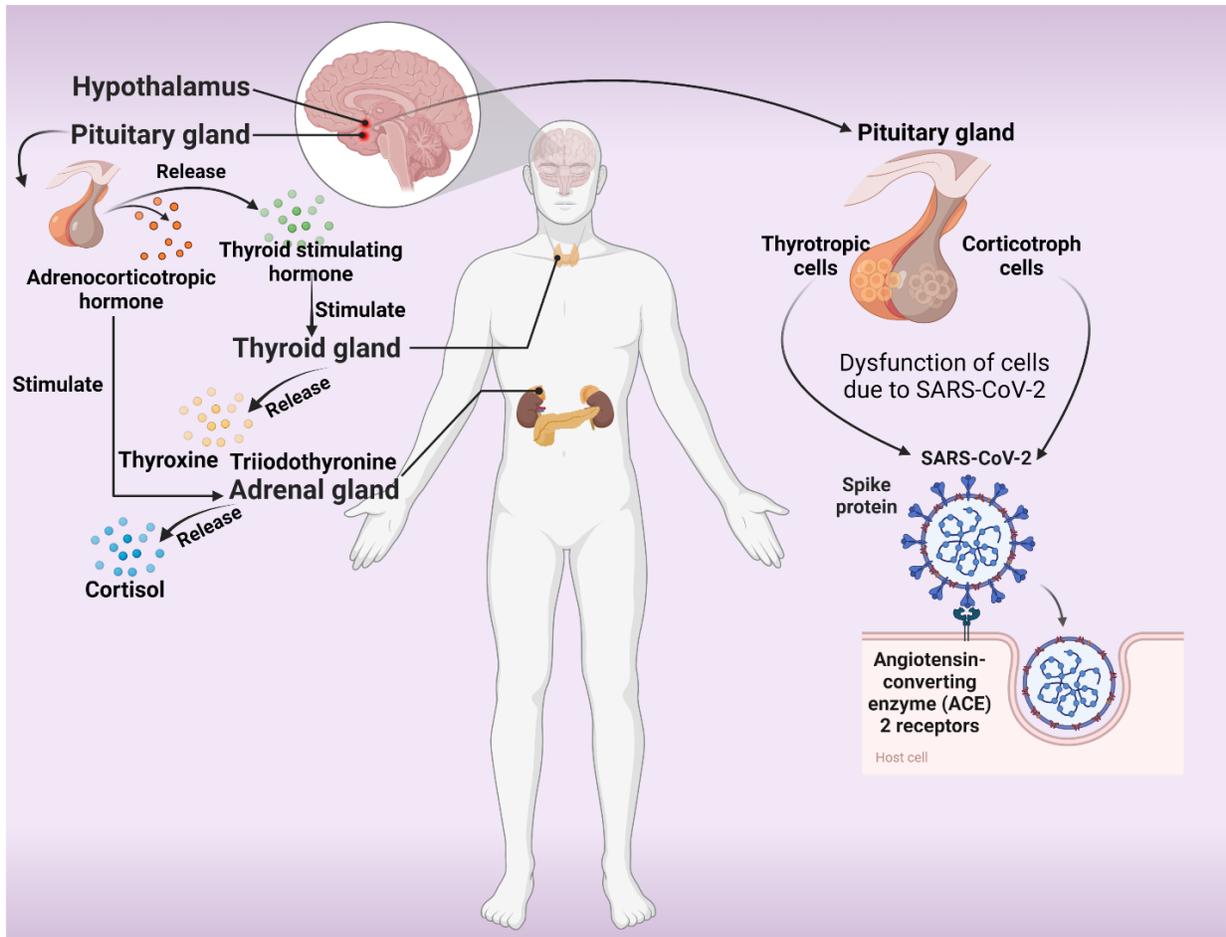

**Figure 9:** SARS-CoV-2, the virus behind COVID-19, impacts the endocrine system, notably the pituitary gland, crucial for hormone regulation. Direct Viral Damage: The virus can infect pituitary cells directly, disrupting their function and hormone production. Inflammatory Response: COVID-19 triggers significant inflammation, potentially interfering with hormone production and release in the pituitary gland. Ischemia and Hypoxia: Severe cases of COVID-19 can lead to reduced blood flow and oxygen levels in organs, including the pituitary gland, damaging cells and impairing hormone secretion. HPA Axis Dysfunction: SARS-CoV-2 may disrupt the hypothalamic-pituitary-adrenal (HPA) axis, crucial for stress response, leading to inadequate cortisol production and adrenal insufficiency. Thyroid Dysfunction: Damage to pituitary thyrotropic cells can disrupt the secretion of thyroid-stimulating hormone (TSH), affecting thyroid hormone production and potentially causing hypothyroidism (generated by using BioRender.com).

### 3.8. Lymphatic system

The lymphatic system is an immune system that defends the body against pathogens, maintains body fluid, absorbs digestive lipids, and eliminates cellular detritus. It comprises lymphatic vessels and lymph nodes that extend throughout the body, draining extra fluid from body tissues and returning it to the circulatory system. About more than eight hundred nymph nodes are found in human body and they are connected to one another by lymph vessels [172]. The lymphatic system is prone to SARS-CoV-2 infection [173] because the lymphatic endothelial cells form the structure of lymphatic vessels and sinuses of lymph nodes and have high numbers of ACE2 receptors on their membrane, which provide the site for SARS-CoV-2 infection [174].

In response to a viral infection of the endothelium, the immune cells such as macrophages produce proinflammatory substances such as tumor necrosis factor- (TNF-α) and interleukin-6 (IL-6) that induce inflammation of lymph nodes [175]. Furthermore, after inflammation and infection of lymph nodes due to the virus, neutrophils are polymorphonuclear phagocytic cells; the first immune cells that reach the lymph nodes secrete extracellular traps, which are net-like structures composed of DNA-histone complexes that trap platelets and increase the expression of the coagulant gene *F13A1* that coagulates the lymph in the lymph vessels and ultimately blocks the vessels (Figure 10). Mediastinal lymphadenopathy (inflamed lymph nodes in the chest) is prevalent in COVID-19 patients, as indicated by Pilechian et al. [176], and it was also revealed that elderly patients expressed more mediastinal lymphadenopathy without any discrimination based on age. Moreover, the lymphocyte (immune cells found in lymph tissues) count was also reduced in patients infected with SARS-CoV-2 [176]. By screening clinical databases, Meyer et al., [177] observed thoracic lymphadenopathy in COVID-19 survivors and non-survivors between 2020 and 2022. In total, 177 female patients were analyzed, and thoracic lymphadenopathy was diagnosed with the diameter of their lymph nodes was greater than ten millimeters (mm) as determined by CT. Approximately 130 patients (73.4 % of the sample) were diagnosed with lymphadenopathy, with a higher mean number of infected lymph nodes in non-survivors than in survivors. In addition, non-survivors have larger lymph nodes (55.9 mm) than survivors (44.1 mm) [177]. Additionally, paranasal sinusitis is observed in a twenty-year-old individual diagnosed with COVID-19 [178] due to the virus-induced obstruction of paranasal lymph vessels [173]. Aging is a major risk factor for delayed immune response in the lymphatic system due to a reduction in the lymphatic vessel diameter that causes waste accumulation in the brain [179], and SARS-CoV-2 infection causes lymphatic vessel and lymph node inflammation and blockage [180]. More details about persistent effects of SARS-CoV-2 on human lymphatic system is described in Table 1.

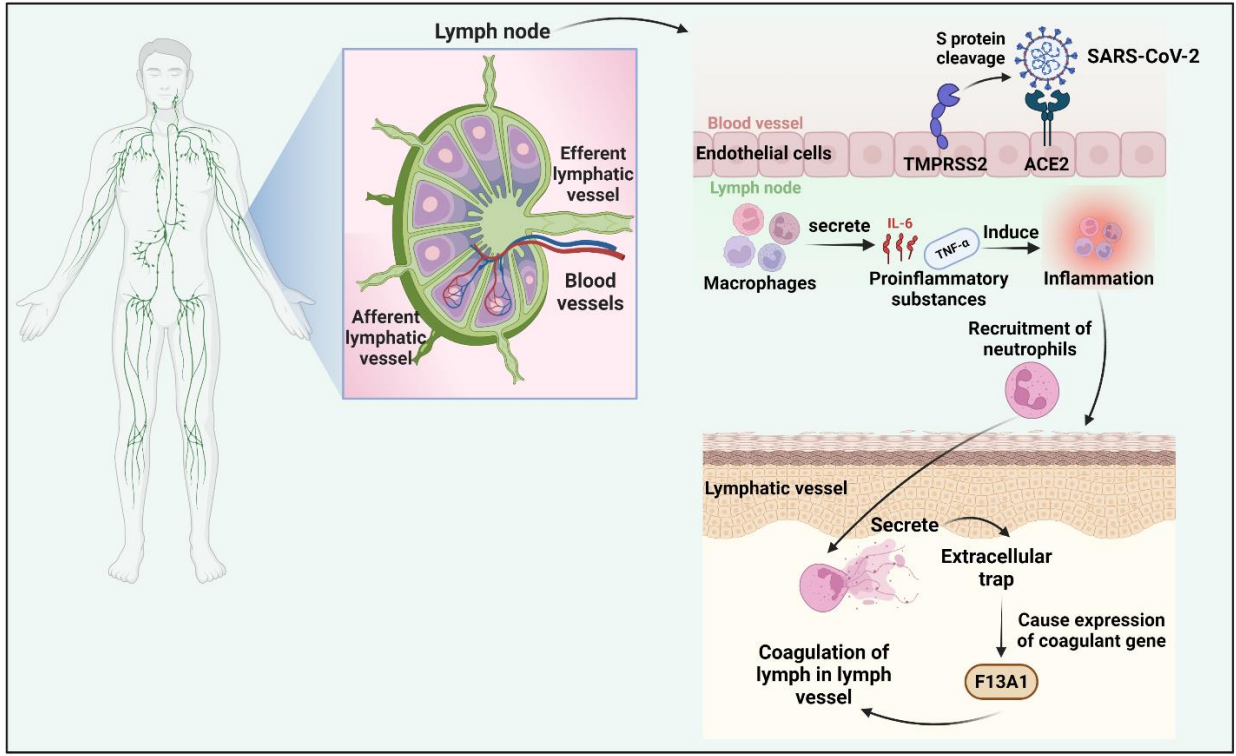

**Figure 10:** SARS-CoV-2 infection can lead to dysfunction in the lymphatic system through several mechanisms: Infection of Lymph Node Endothelium: The virus can infect the endothelial cells lining lymph nodes, triggering inflammation. Macrophage Activation: Infected endothelial cells activate macrophages, which produce inflammatory cytokines like TNF-α and IL-6, contributing to lymph node inflammation. Neutrophil Recruitment and Extracellular Trap Formation: Following lymph node inflammation, neutrophils are attracted to the site. They release extracellular traps, which upregulate the expression of F13A1, a coagulant gene, leading to increased blood clotting within lymph vessels (generated by using BioRender.com).

**Table 1: Long-term effects of COVID-19 on functions of various human systems**

| Country | Study type | Follow up time | Study population | Sample size | Sex % (Male: Female) | Mean age years | Age designations* | Infective human system | Findings | Authors |
|---|---|---|---|---|---|---|---|---|---|---|
| Sweden | Cohort study | 180 days | Hospitalized (48,861 patients) and non-hospitalized (894,121 patients) | N=942,982 | 100:NI | Hospitalized patients=60.6, Non-hospitalized patients=41.4 | Adults | Pulmonary system | Pulmonary embolism, deep venous thromboembolism | [181] |
| Hungary | Cross-sectional study | 180 days | Hospitalized (77 male patients and 95 female patients) | N=172 | 44.77:55.23 | 61 | Older adults | Pulmonary system | Chronic obstructive pulmonary disease, dyspnea | [182] |
| Thailand | Cohort study | 180-300 days | Inpatients and outpatients | N=4,241,201 and 818 patients with severe pneumatic | 41: 59 | Three age groups (18-40, 41-60 and > 60) | Adults and older adults | Pulmonary system | Decreased functional T-cells and pulmonary tuberculosis | [183] |
| United States | Cohort study | NI | NI | N=928 patients | 33 | 53 | Adults | Pulmonary system | The majority of patients (66%) has chronic obstructive pulmonary disease, cystic fibrosis | [184] |
| Switzerland | Longitudinal cohort study | NI | Hospitalized | N=109 | 51:49 | Two age groups (6-11, N=66 and 12-16, N=43) | Children and Adolescents | Pulmonary system | Chest tightness | [185] |
| China | A longitudinal cohort study | 365 days | Non-hospitalized | N=60 | 55:45 | 48 | Adults | Pulmonary system | Chest tightness, Pharyngalgia, chronic bronchitis, tuberculosis | [186] |
| USA | Care team desing (questionnaire) | 180 days | Hospitalized | N=108 | 25:75 | 46 | Adults | Pulmonary system | Chronic obstructive pulmonary disease, pulmonary fibrosis | [187] |
| Israel | Prospectively study | 90.6 days | Out-patients | N=71 | 66:34 | 52.6 | Adults | Pulmonary system | Chronic obstructive pulmonary disease, pulmonary fibrosis, fatigue, dyspnea | [188] |
| Spain | A prospective observational longitudinal study | 365 days | Hospitalized | N=305 | 37.7:62.3 | 87 | Older adults | Pulmonary system | Pulmonary embolism | [189] |
| Italy | Cross-sectional study | <60 days, N=31; 60-119 days, N=30; and >120 days, N=68 | Hospitalized, N=6; Non-hospitalized, N=123 | N=129 | 51.9:48.1 | 11 | Children | Pulmonary system | Chest pain, chest tightness | [158] |
| Mexico | Case report | At birth time | Hospitalized | N=1 | NI:100 | 0.002 | Neonate | Pulmonary system | Reduced oxygen saturation (87%), cyanosis and dyspnea | [190] |
| China | Retrospective study | NI | Out-patients | N=2086 | 52.9:47.1 | 85 | Older adults | Cardiovascular system | Heart failure or stroke, hypertension and coronary artery disease | [191] |

| Israel | Prospectively study | 90.6 days | Out-patients | N=71 | 66:34 | 52.6 | Adults | Cardiovascular system | Ischemic heart disease refers to heart weakening caused by reduced blood flow to heart | [188] |
|---|---|---|---|---|---|---|---|---|---|---|
| Denmark | Cohort study | 195 days | Hospitalized | N=417 | 44.6:55.4 | Four age groups (≥18-40, N=22; 41-60, N=96; 61-80, N=218; and ≥81, N=81) | Adolescents, adults and older adults | Cardiovascular system | Cardiovascular disease | [192] |
| Malta | A case-control study | 173.5 days | Hospitalized | N=249 including 174 cases and 75 controls | 41.36: 58.64 | 46.1± 13.8 | Adults | Cardiovascular system | Cardiac injury, endothelial cells dysfunction, atrial fibrillation, hypertension and hyperlipidemia | [193] |
| Spain | A prospective observational longitudinal study | 365 days | Hospitalized | N=305 | 37.7:62.3 | 87 | Older adults | Cardiovascular system | Congestive heart failure, arterial embolism, atrial fibrillation and deep venous thrombosis | [189] |
| Italy | Cross-sectional study | <60 days, N=31; 60-119 days, N=30; and >120 | Hospitalized, N=6; Non-hospitalized, N=123 | N=129 | 51.9:48.1 | 11 | Children | Cardiovascular system | Myocarditis | [158] |

| | | days, N=68 | | | | | | | | |
|---|---|---|---|---|---|---|---|---|---|---|
| USA | Single-center retrospective study | NI | Hospitalized | N= 27 | 52:48 | 59.8 | Adults | Neurological system | Persistent altered mental status, 74% (N=20) patients with COVID-19 associated encephalopathy, ischemic strokes, diffuse hypoattenuation, subcortical parenchymal hemorrhages, focal hypodense lesions in regions of cerebral vasospasm diagnosed by computed tomography. Moreover, magnetic resonance imaging findings included diffuse involvement of deep white matter, the corpus callosum and the basal ganglia. | [194] |
| Iran | Case study | NI | Hospitalized | N=4 | 50:50 | Four patients have different ages (56, N=1; 24, N=1; 65, N=1; and 71, N=1) | Adults and older adults | Neurological system | Persistent altered mental status-seizure, loss of consciousness | [195] |
| Switzerland | Longitudinal cohort study | NI | Hospitalized | N=109 | 51:49 | Two age groups (6-11, N=66 and 12-16, N=43) | Children and Adolescents | Neurological system | Headache | [185] |
| USA | A case study | 7 days | Hospitalized | N=1 | 100:NI | 0.019 | Infant | Gastrointestinal system | Pharyngitis | [196] |
| USA | A multicenter cohort study | NI | Hospitalized | N=2031 | 57.4: 42.6 | 57.8 | Adults | Gastrointestinal system | Abdominal pain, gastrointestinal hemorrhage and pancreatitis, diarrhea, nausea, vomiting, | [197] |
| Greece | A single-center cross-sectional study | Hospitalized duration is 5.76-7.69 days | Hospitalized | N=161 | 50.93:49.06 | 70.86 | Older adults | Gastrointestinal system | Abdominal pain, gastrointestinal tract bleeding, inflammatory bowel disease, gastrointestinal tract infection with *Clostridium difficile*, dysphagia, diarrhea, gastrointestinal cancer, ischemic colitis and food impaction | [198] |
| Italy | Systematic study | Long-COVID-19 patients or survivors | Hospitalized | COVID-19 long haulers | 100:NI | NI | NI | Reproductive system | Male erectile dysfunction | [199] |

| Poland | An online survey | NI | Hospitalized and Non-hospitalized | N=1644 | NI:100 | 23 | Adults | Reproductive system | Increased risk of sexual dysfunction with decreased libido and lower sexual frequency | [200] |
|---|---|---|---|---|---|---|---|---|---|---|
| Switzerland | A cohort study | 180 days | Non-hospitalized | N=501 | 94:6 | 21 | Adults | Reproductive system | Reduced follicle stimulating hormone, luteinizing hormone and testosterone hormone, sexual dysfunction | [201] |
| Israel | Prospectively study | 90.6 days | Out-patients | N=71 | 66:34 | 52.6 | Adults | Musculoskeletal system | fatigue, muscle weakness and pain | [188] |
| USA | A case study | 14 days | Non-hospitalized | N=1 | NI:100 | 24 | Adult | Musculoskeletal system | Mitochondrial respiration rate in skeletal muscle is reduced, reduction in isokinetic torque during knee extension and knee flexion, muscle strength is also decreased | [202] |
| USA | A case report | Patient lost to follow up after discharge | Hospitalized | N=1 | NI:100 | 44 | Adult | Endocrine system | Thyroid dysfunction, nausea, patient developed adrenal insufficiency and central diabetes insipidus | [203] |
| UAE | A case report | 10 days | Hospitalized | N=1 | 100: NI | 51 | Adult | Endocrine system | Patient developed adrenal insufficiency confirmed by low morning cortisol level | [204] |

| China | | | | | | | | | and cause alteration in endocrine function | |
|-------|-------|-------|-------|-----|------|------|-------------|-------------|-------|-----|
| China | Autopsy study | Patient lost to follow up after discharge | Hospitalized | N=12 | 58.33:41.67 | 70.58 | Older adults | Lymphatic system | Abundant production of inflammatory factors (IL-6, IP-10, TNFα and IL-1β), the spleen and hilar lymph nodes contained lesions with tissue structure disruption and immune cell dysregulation including lymphopenia and macrophage accumulation | [205] |
| China | A case report | 180 days | Hospitalized | N=1 | NI:100 | 34 | Adult | Lymphatic system | Abnormal lymph nodes with thickened lymphatic cortex | [206] |

*Age designations= according to the American Medical Association; Neonates or newborns (birth to 1 month), Infants (1 month to 1 year), Children (1 year through 12 years), Adolescents or teenagers (13 years through 17 years, Adults (18 years or older) and Older adults (65 and older)

*NI= Not informed

## 4. Advanced diagnostic strategies for COVID-19

SARS-CoV-2 detection methods can be broadly categorized into three modalities, including serology tests, nucleic acid detection, imaging-based tests (Figure 11). Serological detection includes antigen and antibody tests which are suitable for rapid testing. The antibody test detects the IgG and IgM antibodies produced by the patient's immune system against SARS-CoV-2. The SARS-CoV-2 antigen tests detect either whole viral particle or spike protein [207].

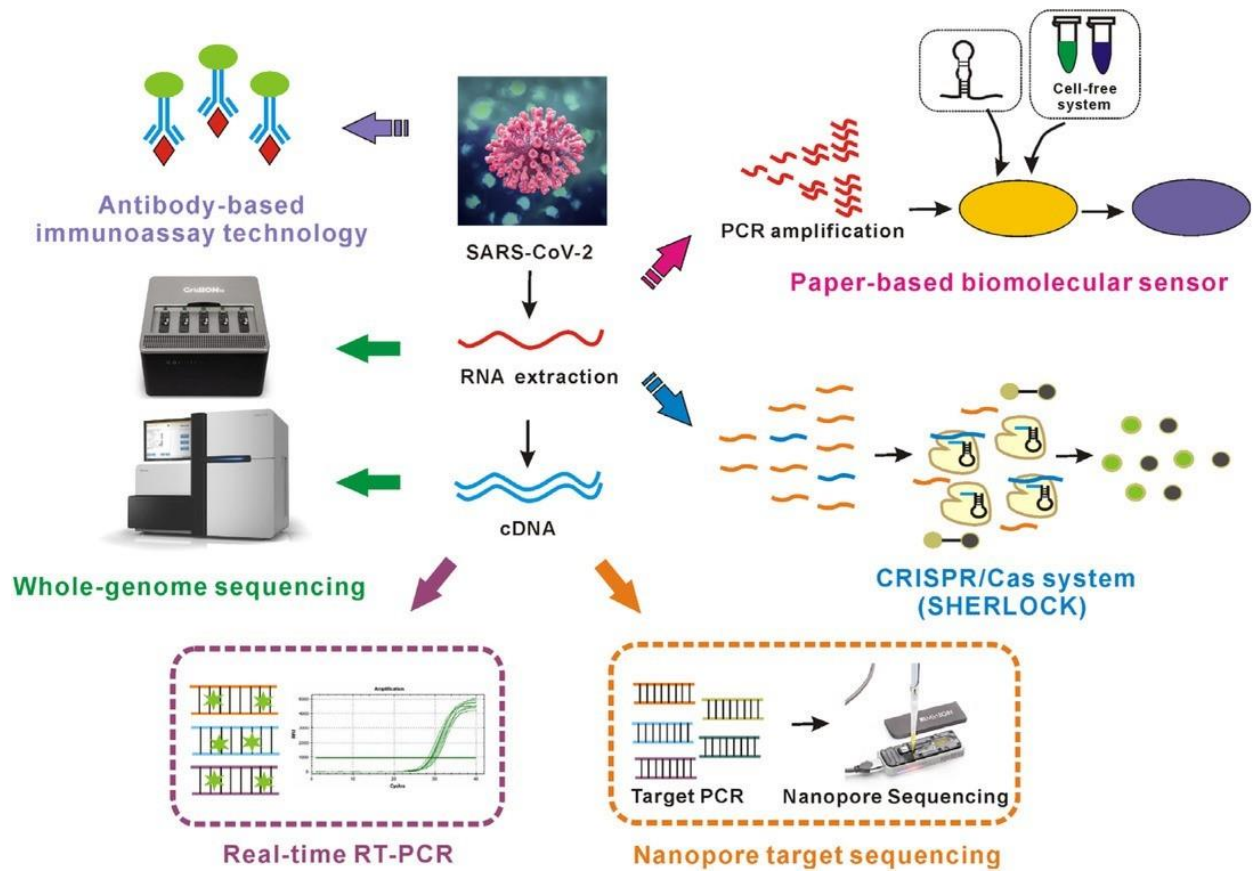

**Figure 11:** State-of-the-art SARS-CoV-2 detection approaches.

A real-time fluorescence reverse transcription PCR tool qualitatively identifies SARS-CoV-2 nucleic acid. The kit contains oligonucleotide primers and dual-labeled hydrolysis probes for the specific detection of SARS-CoV-2 from sections of the SARS-CoV-2 genome's Open Reading Frame 1ab (ORF1ab) gene and nucleocapsid gene (N) [208,209]. Primers and probes for the human RNase P gene (labeled with CY5) are employed as an endogenous internal control. RNA extracted and purified from upper and lower respiratory tract specimens is reverse transcribed to cDNA and amplified in a one-step master mix in an RT-PCR machine. The probes have a reporter on the 5' end and a quencher on the 3' end [209]. During PCR amplification, Taq DNA polymerase with 3-5 exonuclease activity degrades probes attached to amplified templates, separating the reporter dye and quencher, and creating fluorescent signals that rise with each cycle. Based on the signal change, the PCR equipment automatically creates a real-time amplification curve for each optical channel and generates cycle threshold values, which the operator analyzes to detect the presence or absence of SARS-CoV-2 RNA. Although gold standard, diagnostic procedure based on the RT-PCR test is laborious due to sample preparation and quality control measures [210]. The patients with suspected SARS-CoV-2 infections can also be evaluated using chest CT.

CT is a medical imaging technique that utilizes X-rays to generate three-dimensional images of lung parenchymal lesions and other complications, such as pulmonary embolism [211]. Patients are positioned in dorsal recumbency with their forearms behind the head for a thoracic CT scan. The patient maintains ten seconds of profound inspiration during the acquisition. This acquisition must encompass the entire thorax, from the apex to the costodiaphragmatic recess [55]. Nonetheless, CT alone may be insufficient, so some patients with normal radiological findings in the early phases of the disease may be infected with Corona virus. Integrating Artificial intelligence (AI)-based testing procedures, are quick and accurate in diagnosing COVID-19 in patients through imaging using X-rays and computed tomography (CT) [212,213]. Through these advanced technologies, an automatic image is formed with limited contact between laboratory technicians with suspects. After acquiring an image, the image is segmented to detect pulmonary infection, which facilitates quantitative analysis and COVID-19 diagnosis [212]. Moreover, Reichardt et al. [214] viewed the image of the inflammatory infiltrate, predominantly consisting of macrophages, around the intramyocardial vessel through the X-ray phase-contrast tomography.

AI is revolutionizing COVID-19 screening and diagnosis through thermal face detectors, AI-enabled tools for analyzing medical imaging reports, and voice detection platforms [215,216]. The AI algorithms use the dataset of clinical symptoms, exposure history, and laboratory testing to swiftly diagnose patients infected with COVID-19, thereby reducing the workload of medical personnel [217]. Several artificial intelligence (AI) techniques, such as machine learning (ML), TensorFlow, neural networks, and others, assist medical professionals in making accurate decisions using CT scan images, as this method is relatively fast and accurate, with results obtaining in seconds and used to diagnose COVID-19 patients [218]. Furthermore, using ML, the early detection of viruses is enabled by recording user movement, temperature, audio, and video by mobile phone-embedded micro-electro-mechanical system sensors. In the data of 30,000 participants between March and August 2020, 23 % were identified as confirmed COVID-19 patients, with ML identifying positive COVID-19 patients with an accuracy of 79 % [219]. Consequently, the smartphone-based AI algorithm for COVID-19 detection can lead to development of a new, convenient, and efficient diagnosis response to the current global epidemic [220]. Machine learning is advantageous for diagnosing not only acute illnesses but also chronic or long-term effects in patients caused by COVID-19. In this regard, Wang [221] utilized machine learning algorithms to identify four clinical sub-phenotypes of COVID-19 within 30–180 days of infection from the electronic health records (US National Patient-Centered Clinical Research Network, INSIGHT, and OneFlorida+) of patients with SARS-CoV-2 infection. These COVID-19 sub-phenotypes were characterized by cardiac conditions (abnormal heartbeat and heart failure), respiratory problems (breathing abnormalities and pulmonary infections), musculoskeletal (musculoskeletal pain, connective tissue disease, and osteoarthritis), neurological (sleep-wake disorders and headaches), and digestive problems (gastrointestinal problems).

Based on the molecular diagnostic instruments for the early detection of SARS-CoV-2 nucleic acid, clustered regularly interspaced short palindromic repeats (CRISPR) [222,223] and next-generation sequencing (NGS) are highly accurate [224]. CRISPR-Cas is a highly adaptable RNA-guided endonuclease (RGEN)-based nucleic acid editing tool [225,226]. The CRISPR system is simple, efficient, and reliable, with two main classes and six types. The first class comprises of types I, III, and IV effectors, whereas the second class comprises of types II, V, and VI effectors. Type I Cas3 nuclease cuts DNA, whereas type III Cas10 nuclease can cut RNA. Type II Cas9 endonuclease cuts DNA, whereas type V Cas12 cuts DNA. Class II CRISPR systems are extensively employed to diagnose infectious diseases [227–229]. The CRISPR/Cas12a detects SARS-CoV-2 accurately, swiftly (approximately 30 minutes), and with greater specificity and sensitivity than conventional RT-PCR. The CRISPR-based diagnostic methods rely on collateral cleavage activity [208,230]. Once activated, Cas12 effector cleaves all single-stranded (ss) DNA/RNA molecules non-specifically, a phenomenon known as collateral cleavage or trans-cleavage [231]. Harnessing this property, researchers designed fluorescently labeled ssDNA/RNA reporter probes to detect visible bands through the lateral flow assay on a paper strip, thereby confirming the existence of SARS-CoV-2 [232]. The specific high-sensitivity enzymatic reporter unlocking (SHERLOCK) technology employs the activity of the crRNA-Cas13a protein complex to precisely identify RNA molecules and cleave RNA reporter collaterally. Zhang et al. [232] used the CRISPR-Cas-13-based SHERLOCK system to identify two signatures of SARS-CoV-2, namely the S and ORF1b genes, by employing two crRNAs. This technique, like a pregnancy test strip, can provide a visible result within an hour [226]. The SHERLOCK system is sensitive to approximately two attomoles (2 aM) of DNA and RNA, or approximately one cp/l. The average sensitivity of an existing RT-PCR technique is approximately 6.25 cp/l [233]. Moreover, SHERLCOK technology is limited by exponential pre-amplification of the target, which saturates shortly after the reaction begins, making real-time accurate quantification challenging [234].

Next-generation sequencing (NGS) is also a promising diagnostic tool for respiratory, nervous, and gastrointestinal pathogens [235]. This method entails DNA fragmentation, library preparation, massively parallel sequencing, and bioinformatics evaluation. The samples are divided into fragment libraries that are independently amplified and sequenced, generating millions of fragments (small sequences) that can be assembled into a genome readout. Like RT-PCR, this technique provides accurate results within 48 hours [236].

The immunological assay-based diagnostic tool is designed to detect antibodies against SARS-CoV-2 or antigenic proteins in infected individuals [237]. These include the enzyme-linked immunosorbent assay (ELISA), the chemiluminescence immunoassay (CLIA) and the lateral flow immunoassay (LFIA) [238]. These methods rely on the principle of antigen-antibody-specific binding by detecting antibody levels in human serum, and they all have their advantages and disadvantages based on detection efficiency, system cost, ease of operation, and diagnostic limitations [239,240].

Commonly, ELISA is a conventional serological test with an average detection interval between 2 and 8 hours [241]. This method involves coating the virus protein (N, S), S protein subunit (S1), or receptor-binding domain (RBD) on the solid phase carrier of ELISA kits, which then binds with the serum antibody and enzyme-linked antibodies to induce a chromogenic reaction (color formation). Moreover, Varadhachary et al. [242] developed an inexpensive automated ELISA device for detecting SARS-CoV-2-specific immunoglobulin A (IgA) levels (antibody) in human saliva, which can be used to diagnose the disease in less than fifteen minutes.

CLIA operates on a similar principle as ELISA, utilizing the high binding affinity between viral antigens and host antibodies. The CLIA, however, employs a chemical reaction to generate a luminous chemical probe to detect a positive signal. Typically, CLIA provides results within 30 minutes to two hours [243]. Like ELISA, CLIA is a high-throughput assay (tests thousands of samples autonomously) with high accuracy and a low signal-to-noise ratio [237]. Bastos et al. [244] discovered that CLIA detects the SARS-CoV-2 infection earlier than ELISA, with respective sensitivities of 97.8 % and 84.3 %.

The LFIA is an inexpensive, simple, fast, and portable diagnosis tool which is a paper-based detection and analysis platform that rapidly detects virus. This methodology requires three to thirty minutes to complete all process and requires a small quantity of sample such as serum, plasma, whole blood, saliva, tears, urine, and other liquids [245]. The LFIA equipment comprises five elements: a sample pad, a conjugate release pad, a membrane with immobilized antibodies, an adsorbent pad, and an adhesive pad [246]. A drop of whole blood is added to the sample pad and then flows through a conjugate release pad containing S protein-specific immunoglobulin G (IgG) and control IgG (rabbit IgG) conjugated with colored or fluorescent particles. Accompanied by the solution flow, the colored particle-conjugated complexes bind to specific antibodies immobilized in the detection region [247]. In LFIA tests for diagnosing COVID-19, colloidal gold labeled with SARS-CoV-2 antigen is used [248]. This method is simple and convenient, providing instant results, but it is prone to inaccurate results by giving false-negative and false-positive results than other methods [246].

Sensitivity and specificity are two fundamental criteria for evaluating diagnostic strategies; sensitivity refers to the ability of tests to identify positive samples, and specificity refers to the ability of tests to identify negative samples [248]. The fewer false-negative results, the greater the sensitivity and specificity [86]. In COVID-19 diagnosis, the ELISA is less sensitive than the CLIA but more sensitive than the LFIA, indicating that the CLIA-based diagnostic strategy is more effective for diagnosing early infected patients with low antibody concentrations. Furthermore, CLIA has a marginally higher specificity than ELISA and LFIA. In addition, the sensitivity and specificity of the ELISA for IgA detection are extremely high, indicating that IgA may be extensively utilized as a significant indicator for COVID-19 diagnosis [240].

## 5. Advanced treatment strategies for COVID-19

Long-term COVID-19 effects are treated by pharmacological and non-pharmacological techniques [249].

## 5.1. Pharmacological strategy

The pharmacological strategy includes the use of drugs to treat COVID-19 patients, whereas non-chemical intervention is performed to benefit the patients included in the non-pharmacological technique [250,251].

## 5.2. Non-pharmacological strategy

The Rehabilitation is a non-pharmacological strategy may effectively treat individuals with long-term or chronic COVID-19 symptoms by practicing mild aerobic exercise timed according to individual capabilities. Exercise difficulty levels progressively increase within tolerable levels until fatigue and dyspnea improve, usually taking four weeks [252]. Breathing exercises to manage slow, deep breaths and enhance the respiratory muscles' efficiency are also part of rehabilitation. During breathing exercises, the air is inhaled via the nose, expanding the abdominal areas, and exhaled through the mouth. This simple breathing exercise should be done in ten-minutes increments throughout the day [253,254]. Rain et al. [255] analyzed the effectiveness of a 15-day yogic breathing intervention, i.e., short breathing and longer duration breathing techniques, in post-COVID-19 infected patients. Results indicated that both yogic techniques substantially decreased D-dimer, a protein that aids in blood coagulation in veins; consequently, yogic exercise and reduction in D-dimer protein caused a decrease in thrombosis and venous thromboembolism in patients, thereby reducing cardiovascular injury. In comparison, white blood cell counts and respiratory capacity were enhanced. The peak pulmonary oxygen uptake, systematic oxidative stress, and lung ventilation ability are also improved by pulmonary exercise rehabilitation in COVID-19 patients [256]. The cytokine release syndrome is a systematic inflammatory syndrome characterized by many human organ dysfunctions caused by SARS-CoV-2 infection and long-term physical exercise may provide innate immune protection for COVID-19 patients by reducing cytokine release syndrome and improve the response of T cells [257], increase mobilization of natural killer, and CD8 T cells into the blood and tissues [258] which ultimately reduced fever, nausea, fatigue and muscles weakness and pain. Moreover, anti-inflammatory cytokines (IL-4 or IL-10) produced by T cells were also significantly increased after moderate to intensity exercise as well as reduced the expression of inflammatory cytokines such as IL-6, IL-1β and TNF-α, which is caused by COVID-19 [259]. Furthermore, exercise also inhibited myocardial inflammation alongside apoptosis by up-regulating the expression of cytokines such as fibroblast growth factor 21 (FGG21), follistatin-like 1 (FSTL1) and insulin-like growth factor-1 (IGF-1). The immune or lymphatic function is improved by regular exercise of post-COVID-19 patients by reversing pathological changes in vascular endothelial growth factor receptor-3 (VEGFR-3) and Prospero homeobox protein 1 (Prox1) gene expression in lymphatic endothelial cells [260]. The human gastrointestinal system is also prone to SARS-CoV-2 infection, and long-term effects include diarrhea, nausea, vomiting, loss of appetite, and reduced microbial level [261,262]. The intestinal flora can increase the number and function of immunocytes and promote antiviral immunity against SARS-CoV-2 [263]. Therefore, eight weeks of exercise regulated the intestinal microbiota by enhancing the diversity and activity of beneficial micro-organisms [264]. Pilates, music therapy, and neurostimulation, are non-pharmacological strategies for treating long COVID-19 patients. Bagherzadeh-Rahmani et al. [265] conducted a randomized control trial on forty-five participants (24 men and 21 women). Adults with a history of COVID-19 were assigned eight weeks of Pilates training (a form of exercise focusing on balance, posture, strength, and flexibility), and results indicated that training improves pulmonary function by as much as 34 %. In addition, Dehghan Niri et al. [266] found that the Pilates intervention substantially enhanced the musculoskeletal performance of COVID-19 survivors (older adult females) during routine activities. Anxiety harms the recovery and overall well-being of COVID-19 patients; therefore, Giordano et al. [267] conducted a randomized control trial on forty surviving patients in a hospital with music therapy to investigate this non-pharmacological intervention on patients' neurological outcomes. Results indicated that music therapy significantly reduced anxiety and improved oxygen saturation. Tan et al. [234] found that 20 to 30 minutes of music therapy could alleviate a patient's pain and calm their mood, particularly in the vicinity of a hospital, thereby reducing the cardiovascular workload and oxygen consumption to improve physical performance or respiratory muscle strength, as well as having a positive effect on psychological functions. Neuromuscular electrical stimulation is a promising treatment for COVID-19 patients with chronic respiratory and skeletal muscle dysfunction [268]. Applying electrodes to the skin with an electric current modifies the neuromuscular activity of muscles, resulting in enhanced muscle strength and function [269]. Righetti et al. [270] examined the effects of neuromuscular electrical stimulation on muscle mass and functionality in post-COVID-19 patients with persistent cardiovascular complications such as septic shock (a life-threatening condition in which blood pressure drops dangerously low after an infection). The authors hypothesized that this non-pharmacological treatment (neuromuscular electrical stimulation) administered daily for seven consecutive days for forty minutes significantly enhanced muscle strength, including rectus femoris muscle thickness. Therefore, it is speculated that a non-pharmacological strategy is beneficial for improving various human organ dysfunctions.

The pharmacological strategy helps restore the proper function of the human organ system damaged by the SARS-CoV-2 infection [249]. Therefore, Charfeddine et al. [271] evaluated the effect of sulodexide, a novel glycosaminoglycan agent that includes heparin and dermatan sulfate, on 290 patients with long-term COVID-19 symptoms and endothelial dysfunction. The results indicated that the oral doses of sulodexide for three weeks significantly improved endothelial function and reduced thrombosis, inflammation, and chest pain. Furthermore, cardiovascular dysfunction, such as inflammation of cardiomyocytes, is also reduced by using sulodexide. Due to SARS-CoV-2 infection in the long term, irregular heart rhythms (arrhythmia) can cause tachycardia (heart rate over 100 beats a minute), and ivabradine, a hyperpolarization-activated cyclic nucleotide-gated channel blocker, which decreases the heart rate [272]. In a small trial of 34 individuals with persistent tachycardia post-COVID-19, ivabradine was more effective than carvedilol, a beta blocker for patients [272]. Fan et al. [273] performed a study in which the author claimed that patients who had recovered from COVID-19 still had hypercoagulability about a year later. Therefore, apixaban at 5 mg twice a day is very effective as an anticoagulant [274,275]. Furthermore, omega-3 fish oil has multiple benefits for post-COVID-19 patients, as it alleviates fatigue and anosmia (partial or total loss of smell) [276]. Anosmia is one of the persistent symptoms in patients due to COVID-19 due to neurological dysfunction. The SARS-CoV-2 enters the CNS through the olfactory and the brain-blood barrier pathways and invades the brain; therefore, atorvastatin, fampridine, and vortioxetine are pharmacological tools that significantly treat neurological dysfunction in post-COVID-19 patients [277]. Moreover, ultra-micronized palmitoylethanolamide and luteolin, a nutritional supplement, and ivermectin (a nasal spray) have been proven to reduce olfactory dysfunction and improve anosmia [277]. The reproductive dysfunction is also observed in thirty-five years old post-COVID-19 patients admitted to the outpatient clinic with a history of reduced sexual activity and after eighteen months of follow-up with daily use of tadalafil and vitamin D supplementation, improved male penile erectile function [278]. The bismuth subsalicylate is an antidiarrheal agent used for symptomatic treatment of nausea, fever, diarrhea, and other discomforts of the gastrointestinal tract by disrupting SARS-CoV-2 replication, specifically its helicase [279,280]. Hence, pharmaceutical and non-pharmaceutical strategies significantly treat long-term complications in post-COVID-19 patients. More details about effect of pharmaceutical and non-pharmaceutical treatments on different organ systems of post-COVID-19 patients are described in Table 2.

**Table 2: Effectiveness of different pharmaceutical and non-pharmaceutical treatments on human organ system of post-COVID-19 patients**

| Country | Study type | Follow up time | Study population | Sample size | Sex % (Male: Female) | Mean age years | Age designations* | Treatment strategy/Intervention | Targeted human system type | Findings | References |
|---|---|---|---|---|---|---|---|---|---|---|---|
| Indonesia | Case report | NI | Hospitalized | N=1 | NI:100 | 53 | Adults | Non-pharmacological /chest therapy and breathing exercise | Pulmonary system | Chest therapy and breathing exercise helps to maximize lung expansion and improve the patient's breathing effort | [281] |
| Sweden | A quasi-experimental study | 7 days | Non-hospitalized | N=94; intervention group=69.6%, control group=62.1% | 100:NI | For intervention group=50.1, control group=51.5 | Older adults | Non-pharmacological / breathing exercise includes diaphragmatic breathing, deep breathing, huffing (forced expiratory technique) and coughing | Pulmonary system, cardiovascular system | The breathing exercises interventions, even for a short period, effectively improved patients' oxygen saturation, respiratory rate, and heart rate in the intervention group compared to the control group. | [282] |
| Pakistan | A prospective interventional study | 35 days | Hospitalized | N=20; Group 1=N=10; Group 2=N=10 | Group 1=60:40, Group 2=70:30 | Group 1=38, Group 2=41.2 | Adults | Non-pharmacological/Aerobic training and breathing exercise | Cardiovascular system, pulmonary system | Cardiorespiratory fitness, dyspnea, and quality of life significantly improved after five weeks of non-pharmacological treatments of post-COVID-19 patients. | [283] |
| Spain | A quasi-experimental study | 49 days | Hospitalized=44.1%; Non-hospitalized=55.9% | N=68 | 38.2:61.8 | 48.5 | Adults | Non-pharmacological/tele-rehabilitation program based on patient education, physical activity, airway clearing and breathing exercise was structured in eighteen session (three sessions/week) | Pulmonary system, gastrointestinal system, Cardiovascular system | Daily living activities, dyspnea severity, and quality of life improved significantly. Moreover, a significant increase in oxygen saturation before and after six minute' walk patients, heart rate and anosmia, and ageusia are also improved | [284] |
| Brazil | A case report | 5 days | Non-hospitalized | N=1 | 100:NI | 77 | Older adult | Non-pharmacological/active static stretching and physical activity | Pulmonary system, Musculoskeletal system | After five consecutive days of intervention, the oxygen saturation was increased by an average of 1.75%, reaching 89.79% without the help of oxygen therapy, ultimately reducing dyspnea, | [285] |

| | | | | | | | | | | improving muscle health, and increasing mobility | |
|---|---|---|---|---|---|---|---|---|---|---|---|
| India | A case study | 210 days | Non-hospitalized | N=1 | NI:100 | 20 | Adult | Pharmacological/ Ayurveda intervention includes Shadbindu taila marsha nasya for seven days followed by Shadbindu taila pratimarsha nasya (intra nasal oil instillation) for a period of four months, Naradiya laxmivilasa rasa orally for one month; and Non-pharmacological/Traditional Chinese Acupuncture | Neurological system | Integrative approach of Ayurveda and Traditional Chinese Acupuncture cured parosmia within a week and anosmia in four months | [286] |
| India | A case study | NI | Hospitalized and Non-hospitalized | N=11 | 54.55:45.45 | 43.81 | Adult | Non-pharmacological/Yoga Prana Vidya | Pulmonary system | Improvement in bronchial asthma and breathing difficulty | [287] |
| China | A case study | NI | Hospitalized | N=1 | 100:NI | 64 | Older adult | Pharmacological/Bronchoalveolar lavage by using N-acetylcysteine | Pulmonary system | Repeated bronchoalveolar N-acetylcysteine inhalation solution lavage significantly manages of airway | [51] |
| USA | A prospective randomized controlled study | 5 days | Hospitalized | N=97; Standard care alone=50; vagus nerve stimulation=47 | 72.2:27.8 | 58.5 | Adults | Non-pharmacological/Non-invasive vagus nerve stimulation | Pulmonary system | Non-invasive vagus nerve stimulation significantly reduced levels of inflammatory markers, specifically, C-reactive protein and procalcitonin | [288] |
| India | A randomized control trials, observational studies, case series and case reports | NI | Hospitalized and non-hospitalized | An evidence-based review includes various participants | NI | <19 | Infants, neonates, children, Adolescents and adults | Pharmacological/intervention for pain management | Neurological system, gastrointestinal system, pulmonary system, lymphatic system | Acetaminophen and ibuprofen effectively used for pain management. Spasmolytic agents (drotaverine, mebeverine) should be used in case of vomiting, diarrhea and abdominal cramps | [289] |

*Age designations= according to the American Medical Association; Neonates or newborns (birth to 1 month), Infants (1 month to 1 year), Children (1 year through 12 years), Adolescents or teenagers (13 years through 17 years, Adults (18 years or older) and Older adults (65 and older)*NI= Not informed

## 6. Current vaccination strategies for COVID-19

Vaccines are used to prevent and treat COVID-19 due to the high transmission rate of SARS-CoV-2; therefore, currently following types of vaccines are developed for COVID-19 patients.

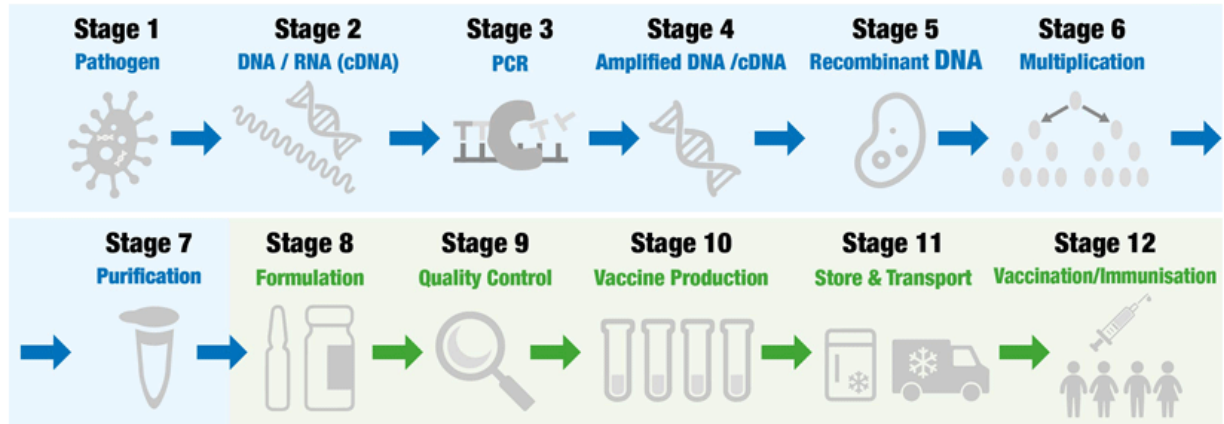

**Figure 12:** Schematic diagram of vaccines production

### 6.1. Inactivated viral vaccines

SARS-CoV-2 infects host cells by binding the S protein to the ACE2 receptor; thus, the S protein is regarded as a vaccine target antigen. Protein-based vaccines include recombinant S proteins, inactivated viruses, and virus-like particle [290]. Inactivated vaccines are produced through the growth of SARS-CoV-2 in cell culture and subsequent inactivation of the virus through several methods, such as the use of beta-propiolactone, formalin, formaldehyde, glutaraldehyde, ultraviolet, and gamma rays [291], without significantly affecting their antigenicity, thus, whole virus particles or parts of the virus cannot use for vaccines. These treatments (beta-propiolactone, formalin, formaldehyde, glutaraldehyde, ultraviolet, and gamma rays) result in viral protein fixation or genome damage. In addition, aluminum hydroxide adjuvants are typically added to this type of vaccine to enhance the immune response it elicits [292]. Using beta-propiolactone, Gao et al. [293] deactivated SARS-CoV-2 and examined its effectiveness in mice, rats, and non-human primate models. According to the results, vaccine induced anti-S and anti-receptor binding domain antibodies and immunized animal serum exhibited neutralizing activity. Various inactivated SARS-CoV-2 vaccines, including Covaxin, KoviVac, Turkovac, FAKHRAVAC, QazVac, KCONVAC, COVIran Barekat, Covilo, CoviVac, CoronaVac, Sinopharm, Sinovac, Sinopharm-Wuhan, VLA2001 and Bharat Biotech BBV 152, have been approved and are in widespread use around the globe [294]. The inactivated vaccines are administered intramuscularly and stored between 2 and 8 °C [295].

Vaccination reduced the long-term effects of SARS-CoV-2 on human organ systems to the point that Budhiraja et al. [296] conducted a two-year observational study to investigate the effect of Covaxin, an inactivated viral vaccine, on post-COVID-19 in India. The result indicated that 90 % of patients acquired fatigue, insomnia, and myalgia symptoms at the time of vaccination, but after one month, about ¾ of the COVID-19-vaccinated patients showed the same symptoms. At four weeks, only 15 % of patients reported the chronic effects of COVID-19. After one year, only 11 % of patients showed symptoms. Furthermore, Gao et al. [297] also claimed the effect of inactivated vaccines in reducing the risk of long-term effects of COVID-19 on respiratory and cardiovascular neurological systems, such as anxiety or depression, chest or throat pain, fatigue, hair loss, headache or migraine, kidney diseases, loss of concentration, loss of smell, loss of taste, myalgia, nausea or vomiting, respiratory symptoms, sleeping disorders, irregular heartbeats, and weight loss. These results align with those of Arjun et al. [298], who investigated the effect of two doses of Covaxin on patients who suffered long-term complications due to post-COVID-19. Aging is negatively associated with the immunogenicity of inactivated viral vaccination in both men and females. As a result, Medeiros et al. [299] compared the T cell and antibody responses of CoronaVac-vaccinated persons (N=101) to those of convalescent patients (recovered from COVID-19) (N=72). The data demonstrated that around 96 % of the vaccinated group had a T cell or antibody response to SARS-CoV-2, compared to 98.5 % of the convalescent group. Furthermore, 55 years old vaccinated people of both sexes generated considerably decreased anti-RBD, anti-nucleocapsid protein IgG, IL-2 production by T cells, and neutralizing antibodies against SARS-CoV-2 as compared to younger individuals of both sexes (<55 years old). Therefore, to increase immune response in older persons, one extra booster dose of CoronaVac is required after two doses of this vaccine to induce long-term protection against COVID-19. Mild side effects of inactivated viral vaccine (Bharat Biotech BBV 152) were observed, such as pain in the injection site, lethargy, fever, muscular soreness, and fever in healthcare adult workers of both sexes in Birjand City, Iran [300]. More detail regarding the effects of inactivated viral vaccines on individuals of different genders is provided in Table 3.

### 6.2. Live attenuated viral vaccine

Live-attenuated vaccines differ from inactivated ones in that the virus remains alive but has been attenuated or altered so that it cannot cause disease but evoke a robust immune response. Because they are similar to natural infections, live attenuated viral vaccines elicit a more substantial and long-lasting immune response. Typically, one or two doses of live, attenuated viral vaccines are sufficient for lifelong protection against disease. However, in case of SARS-CoV-2, the vaccination strategy utilizing live-attenuated vaccines is still being evaluated and could be implemented annually [301]. The immunological mechanism of live-attenuated vaccines typically involves a broad immune response consisting of CD4+ and CD8+ T lymphocytes (T-cells) and antibodies against SARS-CoV-2 (produced by B-cells) and provides long-lasting immunity against the virus [302, 303]. Moreover, to immunize against the new variants of SARS-CoV-2, attenuated viruses could be generated from previously approved live-attenuated vaccines with the appropriate S protein gene inserted from the new mutant strain to generate an updated live-attenuated vaccine that expresses the S protein mutations, which is feasible, as noted by [304]. Because attenuated viruses can revert to their pathogenic forms and cause symptomatic problems at the application site, typically the nasal cavity, this vaccination strategy requires continuous monitoring during development and application [305]. Genetically immunocompromised individuals or those infected with the human immunodeficiency virus (HIV) should not receive live-attenuated vaccination [298] . Moreover, these already afflicted individuals should avoid coming into contact with live-attenuated vaccinated individuals, as their possibility of getting the attenuated virus will increase, which could spread unchecked under a weakened immune system, resulting in severe health issues [299] . Before administering live-attenuated virus vaccines in the community, older individuals should be immunized with non-proliferative vaccines to develop mucosal immunity against SARS-CoV-2. Three live-attenuated virus vaccines, favipiravir, remdesivir, and ribavirin, have been approved and are widely used worldwide [306]. Live-attenuated viral vaccines hinder SARS-CoV-2 from causing persistent effects on human organ systems; therefore, Chen et al. [36] prepared a live-attenuated viral vaccine (dNS1-RBD), applied intranasally on the animal model (mice), for prolonged investigation (one year) on pathological effects caused by SARS-CoV-2 on the respiratory system. The results indicated that dNS1-RBD triggers the immunity response in the vaccinated-infected mice by upregulating the expression of the tissue-resident memory marker CD69, which promotes long-term resistance against the virus after six months of vaccination. Furthermore, dNS1-RBD also generated CD4+ and CD8+ T lymphocytes in the bronchoalveolar space and lungs even after twelve months of vaccination, and no weight loss was observed in mice. Moreover, Mehla et al. [307] aimed to protect infants (male and female) (below six months of age) from long-term effects of COVID-19; therefore, a new candidate of live attenuated viral vaccine, dCOV, was investigated on mice via intranasal or intramuscularly and results indicated that this vaccine induced high levels of neutralizing antibodies without generating any harmful effects in hamsters therefore, it is effective and safe novel vaccine for infants. Table 3 provides additional information on the effects of live attenuated viral vaccines on individuals of various genders.

### 6.3. Viral-vectored vaccines

Rather than possessing antigens by themselves, viral vector vaccines use the translation apparatus of host cells to produce antigens [294]. Therefore, modified viruses (vectors) are used to deliver antigen-encoding genes; in the case of SARS-CoV-2, the gene encoding the virus's surface S proteins is delivered into human cells by other modified viruses. There are two types of viral-vectored vaccines: replicating and non-replicating. After entering the cell, replicating viral-vectored vaccines produce both whole viral particles and the vaccine antigen (SARS-CoV-2 S protein), whereas non-replicating viral-vectored vaccines are incapable of producing whole virus particles and can only produce the vaccine antigen [294]. Currently six non-replicating viral-vectored vaccines available, including Ad5-nCoV, Ad26.COV2. S, ChAdOx1 nCoV-19, Covishield, Gam-Covid-Vac, and Sputnik Light. Comparatively, replicating viral-vectored vaccine candidates are currently undergoing preclinical or clinical testing [308]. Concerning the ChAdOx1 nCoV-19 vaccine, the chimpanzee adenovirus ChAdOx1 carries the gene of SARS-CoV-2 S protein and enters into the nucleus, where DNA polymerase transcribes it into mRNA. mRNA then exits the nucleus and enters the cell plasma, where it attaches to ribosomes to synthesize antigen proteins, which are expressed and released, followed by antigen presentation, which stimulates the immune system. The immunity developed in humans against SARS-CoV-2 depends upon antigen presentation, which allows T cells to recognize antigenic epitopes on the antigen-presenting cell's surface [309]. T lymphocytes then eliminate virus-infected cells, while cell-mediated immunity stimulates macrophages and natural killer cells to eliminate intracellular pathogens [308]. Moreover, 28 days after vaccination with the ChAdOx1 vaccine, the lymphocytes' (CD4+ and CD8+) expression of the immune response regulator genes CD69 and Ki-67 increased [310]. Compared to other vaccine types, viral-vectored vaccines are more stable and require less stringent storage and handling conditions, such that Ad26COV2.S, an adenoviral vector-based vaccine, can be stored at 2 to 8 °C for six months, thereby increasing the global affordability and availability of vaccines [311].

The effect of the ChAdOx1 nCoV-19 vaccine in the reduction of the long-term effect of SARS-CoV-2 in post-COVID-19 patients was investigated by Scherlinger et al. [312], who conducted a nationwide online study among adult patients having symptoms persisting over four weeks following a confirmed COVID-19. The six hundred and twenty questionnaires were completed from respondents having a median age of 44 years, including 83.4 % women. The results indicated that among 380 patients with persistent symptoms including fever, fatigue, brain fog, headache, sleeping issue, costal pain, dyspnea, cough, palpitations, muscle aches, joint pain, anosmia, ageusia and vomiting at the time of vaccination with 21.8 % patients reported with improvement by reducing the persistent symptoms. Furthermore, Strain et al. [313] also claimed to reduce long-term symptoms such as fatigue, brain fog, myalgia, gastrointestinal symptoms, cardiovascular symptoms, and autonomic dysfunction induced by SARS-CoV-2 in post-COVID-19 patients. A prospective observational study conducted by Arnold et al. [314] based on the safer action of the viral-vectored vaccine in patients with long COVID-19 revealed that the ChAdOx1 nCoV-19 vaccine is not associated with worsening of persistent symptoms in post-COVID-19 patients due to SARS-CovV-2.

The viral vector vaccine's reactogenicity (adverse reaction of vaccine) is associated with the patient's gender and age [315]. Therefore, Nachtigall et al. [316] investigated the reactogenicity of the ChAdOx1 nCoV-19 vaccine in healthcare workers of various gender and age who had at least received one dose of the vaccine and results indicated that young women showed highest rates of reactogenicity (swelling or pain at injection site) after a COVID-19 vaccination. More effects of viral-vectored vaccines on individuals of various genders are provided in Table 3.

### 6.4. Protein subunit vaccines

The protein subunit vaccines contain purified antigenic parts of the desired virus rather than a whole virus to trigger an immune response. The protein subunit vaccines include polysaccharides and conjugates [294]. Many protein subunit vaccines, including Zifivax, Noora vaccine, Corbevax, Abdala, Soberana, Soberana Plus, V-01, MVC-COV1901, Nuvaxovid, IndoVac, Razi Cov Pars, VidPrevtyn Beta, COVOVAX, SKYCovione, TAK-019, SpikeGen, Aurora-CoV, and EpiVacCorona, are approved worldwide [317]. The immunogenicity of Soberana and Soberana Plus was evaluated in post-COVID-19 patients by Ramezani et al. [318] and revealed that patients generated anti-Spike lGg antibodies even after six months of vaccination to protect against any long-term effects caused by SARS-CoV-2. Moreover, protein subunit vaccines cause prolonged antibody generation when used as a booster vaccine. For example, ZF2001, a protein vaccine, was used as a booster vaccine following an inactivated viral vaccine (Sinopharm) and could increase the neutralizing antibodies (CA1 and CB6) [319] that provides long-term protection against the delta or Omicron variant of SARS-CoV-2 [232]. Sadat Larijani et al. [320] investigated the long-term effects (18 months) of protein subunit vaccines on post-COVID-19 patients and revealed that two doses of PastoCovac and one dose of PastoCovac Plus, a protein vaccine, improved the gastrointestinal dysfunction and Canker sores in patients. Regarding harmful effects of protein subunit vaccine, fever is a possible side effect of the Novavax COVID-10 protein subunit vaccine in pregnant adult women and high fever increase the chances of birth defects [321]. Table 3 provides additional information on the effects of protein vaccines on individuals of various genders.

### 6.5. mRNA vaccines

Nucleic acid vaccines containing antigens encoded by DNA or RNA are transmitted by viral vectors such as adenoviruses; these vaccines solve problems caused by more conventional vaccines, such as the risk of reversion to virulence in live-attenuated vaccines and the requirement for additional adjuvants [322]. Nucleic acid vaccines are also very effective because they replicate a live, in-situ infection by expressing antigens after immunization; this primes both B and T cell responses and develops an adaptive immune response directed towards the encoded target antigen [323].

Generally, two classes of mRNAs, Non-replicating and self-amplifying mRNA, are used as vaccine genetic vectors. Both utilize the host cell translational machinery for the production of the antigen target and launch of an adaptive immune response; non-replicating-mRNA only encodes protein antigen of interest, whereas self-amplifying mRNA is also capable of encoding proteins, permitting RNA replication [324]. Currently, COVID-19 mRNA vaccines are non-replicating vaccines [325]. WHO documented 337 COVID-19 vaccine candidates have been under development since February 8, 2022, with 47 mRNA vaccines and 23 clinical trials currently underway [326]. The development of the Pfizer-BioNTech, Moderna, and CureVac vaccines was the quickest in medical history [327].

The mRNA vaccines improve the adverse long-term effects of SARS-CoV-2 in various human organs [328]. Ayoubkhani et al. [328] conducted a community-based cohort study to estimate the association between Pfizer-BioNTech and Moderna vaccinations and long COVID-19 symptoms in adults before and after vaccination. The results indicated that after the first dose of the vaccine, the symptoms such as loss of smell, loss of taste, and trouble sleeping were reduced. After the second dose of both vaccines, fatigue, headache, and trouble sleeping symptoms decreased. In addition, Lucas et al. [329] investigated the immune response of COVID-19 recover and naive individuals who received the mRNA vaccine against SARS-CoV-2. After six months of vaccination, plasma samples from post-COVID-19 patients demonstrated greater neutralizing activity than uninfected donors [330]. Due to neutralizing antibodies production in post-COVID-19 patients by mRNA vaccine, the long-term effects of SARS-CoV-2 on various organ systems such as the cardiovascular system (tachycardia, heart failure, myocardial infarction, and deep vein thrombosis), gastrointestinal system (constipation and esophageal disorders), kidney (chronic kidney disease), neurological system (seizure and stroke), musculoskeletal system (joint and muscle pain), pulmonary system (hypoxemia and shortness of breath), reproductive organs, lymphatic system, endocrine system [331]. Regarding the safety and efficacy of mRNA vaccines in children aged 5–11 years, Watanabe et al. [332] performed a meta-analysis of vaccinated children. They found that this form of vaccine has no adverse effects on minors. In addition, Shimabukuro et al. [333] investigated the impact of mRNA vaccines on pregnant women aged sixteen to fifty-five years in the United States. The results revealed that injection site pain was reported as a side effect of the vaccine and that among 3958 participants, 86.1 % of pregnant women had live births, 13.9 % lost their pregnancy, and there were no neonatal deaths. Table 3 provides additional information on the effects of mRNA vaccines on individuals of various genders.

### 6.6. DNA vaccines

DNA vaccines deliver their genes or gene fragments, encoding immunogenic antigens, to the host's cells by employing DNA plasmids as a vector and eliciting an effective immune response. The vaccine is formulated so that the genetic material is translocated to the nucleus of the host cell, and after reaching the nucleus, the mammalian promoter in the vector is activated, triggering the transcription of the gene used for the vaccine by the host's cellular machinery [334]. Antigen-presenting cells are the primary recipient of genetic material (genes) translated into protein, which binds to major histocompatibility complex (MHC-) I or II through peptides. Myocytes (Cells other than antigen-presenting cells) use MHC-I for antigen presentation. In contrast, dendritic cells (antigen-presenting cells) use MHC-II, resulting in presentation of antigens to both CD4+ and CD8+ T cells, which produce an immune response [335]. Numerous DNA vaccines, including INO-4800, AGO301, AGO302, GLS-5310, GX-19 N, and COVID-eVax, are undergoing phase II and III clinical trials, while only ZyCoV-D has been authorized for use [336,337]. Vaccine storage is a crucial factor in vaccinology; therefore, refrigerated storage (below 2 °C) is required to preserve the content of live vaccines [301]. In contrast, DNA vaccines are highly stable and require less refrigeration (2 to 8 °C) [338].

DNA vaccines provide long-term protection against SARS-CoV-2 in post-COVID-19 patients; therefore, Ahn et al. [339] conducted a non-randomized phase 1 trial for 52 weeks to investigate the long-term effects of two DNA vaccines, GX-19 and GX-19N, in post-COVID-19 patients aged 19–54 years who received two 3 mg doses intramuscularly using an electroporator. Results indicated that patients who received the GX-19 vaccine displayed more severe long-term effects (fatigue: 15 % and neurological disorders: 20 %) than those who received the GX-19N vaccine (no fatigue and fewer nervous system disorders: 10 %). Therefore, the authors claimed that GX-19N induces humoral and broad SARS-CoV-2-specific T-cell response and was selected as a vaccine candidate for phase II immunogenicity trials. Furthermore, Babuadze et al. [340] designed two DNA plasmid vaccines, plDV-V1 and plDV-V5, to assess their effects on various organs of non-human primate models and results indicated that both vaccines protect animal weight loss in long COVID-19 with reducing pulmonary and cardiovascular dysfunctions. DNA vaccine improves the functionality of various organs in non-human primate models and induces immunity against long-term effects due to SARS-CoV-2 in post-COVID-19 patients. Therefore, Khobragade et al. [341] investigated the three doses of ZyCoV-D vaccine (2 mg per dose) applied intradermally via a needle-free injection system 28 days later in post-COVID-19 patients aged 36.4 years with long-term symptoms such as hypertension, diabetes, obesity, chronic pulmonary diseases, chronic kidney disease or chronic heart disease and revealed that the two doses of this vaccine improved the adverse effects of SARS-CoV-2 (chronic heart, lung, and liver diseases became stabled and diabetes of patients was also controlled) with 100 % efficiency. Concerning the safety and efficacy of the ZyCoV-D vaccine in male adults aged 18 to 55 years, Momin et al. [342] conducted a phase 1 clinical trial. They discovered that this vaccine is risk-free for adults and does not cause any adverse reactions; however, most adverse reactions, such as fever, are reported in patients over 55 years [295,343]. Table 3 provides additional information on the effects of DNA vaccines on individuals of various genders.

### 6.7. Nanoparticle vaccines

Nanoparticle-based vaccines deliver SARS-CoV-2 antigens, enhancing vaccination outcomes against COVID-19 [344]. Nanoparticles are nanoscale-sized structures that mimic the structural features of natural viruses, making them the most promising for the development of a next-generation vaccine against SARS-CoV-2 that elicits a strong neutralizing antibody response or stronger antibody-based immunity. Approximately 26 nanoparticle-based vaccine candidates, including lipid nanoparticles, virus-like particles, and protein nanoparticles, undergo human clinical trials, while an additional sixty vaccines are in various phases of preclinical development [345]. The primary route of nanoparticle vaccines is intramuscular because this route ensures efficient bio-distribution, which contribute overall immunogenicity. After intramuscular administration of a nanoparticle vaccine to humans, the vaccine enters lymphatic vessels and is transported to lymph nodes. Furthermore, tissue-resident antigen-presenting cells (dendritic cells) and muscle cells ingest nanoparticle vaccinations [346]. The primary goal of nanoparticle vaccines is the efficient delivery and expression of antigens within the cell, their processing, uploading of epitopes on the MHC, and exposure on the antigen-presenting cell surface for T-cell activation, which induces an immune response by producing B-cell antibody [347]. The immunogenicity of ferritin nanoparticle-based SARS-CoV-2-receptor-binding domain vaccine was evaluated in a non-human primate model by Wang et al. [168] and revealed that vaccine-induced antibody response in animals even after seven months of vaccination to protect against any long-term effects caused by SARS-CoV-2. Furthermore, Russell et al. [348] reported that COVID-19 survivors persistently experienced pulmonary embolisms, ischaemic stroke, myocardial infarctions, and cardiovascular diseases as long-term consequences of SARS-CoV-19, and a nanoparticle-based vaccine is a promising strategy for improving thrombolytic treatment in COVID-19 survivors by delivering thrombolytics with nanoparticles. Gender is associated with the immunogenicity of nanoparticle vaccination, and males and females show different responses to vaccines. Therefore, Vulpis et al. [349] exposed the lipid nanoparticles to the whole blood of eighteen healthy donors, including ten females and eight males, and the results indicated that males and females generated significant differences in natural killer production. Moreover, ageing causes the production of fewer natural killers in older adults than in younger males and females. Regarding side effects, the lipid nanoparticles can cause acute pain at the injection site, swelling, and fever [350]. Moreover, table 3 provides additional information on the effects of nanoparticle vaccines on individuals of various genders.

**Table 3: Effect of various vaccines on post-COVID-19 individuals**

| Country | Study type | Follow up time | Study population | Sample size | Sex % (Male: Female) | Mean age years | Age designations * | Vaccine name | Vaccine type | Dose (s) | Findings | Reference |
|---|---|---|---|---|---|---|---|---|---|---|---|---|
| UAE | Cross-sectional survey | 21 days | Non-hospitalized | N=1080 | 29.6:70.4 | 37.22±13 | Adult | Sinopharm | Inactivated viral vaccine | 2 | Fatigue and injection site pain are the most common side effects in females aged less than 49 years. | [351] |
| Sweden | Case-control study | NI | Hospitalized | N=5905 | 13.63:86.37 | 50 | Adult | Measles-mumps-rubella vaccine | Live viral vaccine | 2 | A significant effectiveness of about 57% at preventing symptomatic disease in men | [352] |
| Germany | Survey-based study | > 60 days | Non-hospitalized | N=599 | 27.7:72.3 | 39 | Adult | ChAdOx1 nCoV-19 | Viral-vectored vaccine | 1 | The females of younger age showed more side effects, such as injection site pain, than the males. | [353] |
| China | Large scale, double blind, randomized placebo-controlled phase 2 trial | 57 days | Hospitalized | N=3295 | 56.3:43.7 | 45 | Adult | MVC-COV1901 | Protein subunit vaccine | 2 | Injection site pain, malaise, and fever are side effects of this vaccine in both sexes. | [354] |
| Germany | Survey-based study | > 60 days | Non-hospitalized | N=599 | 27.7:72.3 | 39 | Adult | Pfizer-BioNTech and Moderna | mRNA vaccine | 2 | Due to these mRNA vaccines, younger females showed headache and fatigue side effects. | [353] |
| USA | Multi-center clinical trial | 365 days | Non-hospitalized | N=20 | 55:45 | 34.5 | Adult | INO-4800 | DNA vaccine | 2 | No adverse side effects was observed in vaccinated individuals | [355] |

| Germany | Prospective interventional study | NI | NI | N=18; cohort 1 (2 µg dose), cohort 2 (4 µg dose) | Cohort 1 (63:38), Cohort 2 (75:25) | Cohort 1 (39.6), cohort 2 (39.1) | Adult | Lipid-nanoparticle vaccine (CVnCoV) | Nanoparticle vaccine | 2 | This vaccine boosts the immune response even at a low dose in vaccinated individuals (males and females). | [356] |

*Age designations= according to the American Medical Association; Neonates or newborns (birth to 1 month), Infants (1 month to 1 year), Children (1 year through 12 years), Adolescents or teenagers (13 years through 17 years, Adults (18 years or older) and Older adults (65 and older)

*NI= Not informed

## 6.8. Comparative effectiveness of various vaccine types on post-COVID-19 individuals

Vaccination against SARS-CoV-2 reduces the acute and chronic effects of COVID-19. Nisar et al. [357] conducted test-negative case-control studies in Pakistan to evaluate the efficacy of diverse COVID-19 vaccines, including mRNA, inactivated, and viral vector vaccines. Phone calls were made to adult residents of Karachi, Pakistan, who tested positive for COVID-19 using RT-PCR at the Aga Khan University Hospital diagnostic facility between June and September 2021. One thousand five hundred ninety-seven cases and 1590 controls were enrolled; 38.1 % of cases and 53.3 % of controls were fully immunized. Those who received a two-dose vaccination regimen had an average gap of 27,3 days between the last vaccination dose and the RT-PCR test. Sinopharm was the most frequently utilized vaccine (61.6 %), followed by Sinovac (25.6 %) and CanSinoBio (8.9 %). Whereas only 2.3 % (33) individuals have received Sputnik-V, eight participants (0.5 % each) received Pfizer and Moderna's mRNA vaccines; six participants (0.4 %); AstraZeneca received 0.4 %; and only two participants (0.1 % had received Janssen. Two doses of mRNA vaccines (Pfizer and Moderna) demonstrated the highest efficacy (67.4 %), followed by Spuntnik-V (58.5 %), Sinovac (49.3 %), and Sinopharm (33.8 %). In contrast, a single dose of CanSinoBIO and AstraZeneca were 47.9 % and 31.8 % effective against the delta variant of SARS-CoV-2, respectively [357]. In addition, inactivated COVID-19 vaccines were only moderately effective against persistent infection with COVID-19 in Pakistani adults, and a supplemental dose is required to increase their efficacy, whereas mRNA vaccines were more effective. More detail regarding comparative effectiveness of various vaccines is described in Table 4.

**Table 4: Comparative effectiveness of various vaccine types on post-COVID-19 individuals**

| Follow up time | Sex % (Male: Female) | Mean age | Age * | Variant | Inactivated viral vaccine | Live attenuated viral vaccine | Viral-vectored vaccine | Subunit vaccine | mRNA vaccine | DNA vaccine | Nanoparticle vaccine | Findings | References |
|---|---|---|---|---|---|---|---|---|---|---|---|---|---|
| 77 days | 95.8:4.2 per recipient group | 70 | Older adults | Delta and Omicron variants | NI | NI | NI | NI | BNT162b2/ 30 µg for $1^{st}$, $2^{nd}$ and $3^{rd}$ dose; mRNA-1273/50 µg for $1^{st}$ and $2^{nd}$ dose and 100 µg for $3^{rd}$ dose | NI | NI | Third dose of each vaccine significantly reduced COVID-19 outcomes in patients, but mRNA-1273 is more effective than BNT162b2. | [358] |
| 300 days | 48.62:51.37 | ≥ 18 | Adults | Alpha and Delta variants | NI | NI | ChAdOx1 n CoV-19/two doses | NI | BNT162b2/ two doses | NI | NI | Both vaccines significantly reduced COVID-19 outcomes in patients, but BNT162b2 vaccine appeared more efficacious than the ChAdOx1 n CoV-19 | [359] |
| 15-175 days | NI | 45-75 | Adults and older adults | Delta variants | NI | NI | ChAdOx1 n CoV-19/three doses | NI | BNT162b2/ three doses; mRNA 1273/three doses | NI | NI | Two doses of all vaccines offered high effectiveness (more than 90%). In contrast, three doses of each vaccine increased effectiveness nearly to 100% in individuals of various ages, and it was concluded that individuals should be vaccinated with at least two doses to reduce the long-term effects of COVID-19. | [360] |
| 107 days | 52.4:47.6 | 49.3 | Adults | Delta and omicron variants | NI | NI | NI | NI | BNT162b2/ three doses; mRNA 1273/three doses | NI | NI | Both vaccines significantly reduced COVID-19 outcomes in patients, but mRNA 1273vaccine appeared more efficacious than the BNT162b2. | [361] |
| NI | NI | NI | Adults | Delta and omicron variants | NI | NI | ChAdOx1 n CoV-19/two doses | NI | BNT162b2/ two doses; mRNA 1273/two doses | NI | NI | The vaccine effectiveness against SARS-CoV-19 was 71% in the first dose of vaccines and 87% in the second dose of vaccines | [362] |
| 168 days | 92.7:7.3 | 69 | Older adults | Alpha and delta variant | NI | NI | NI | NI | BNT162b2/ three doses; mRNA | NI | NI | After a 24-week assessment, the third dose of each vaccine significantly reduced COVID-19 outcomes in patients, but mRNA- | [363] |
| | | | | | | | | | 1273/three doses | | | 1273 is more effective than BNT162b2 | |
| NI | NI | ≥ 18 | Adults | Alpha and delta variant | NI | NI | NI | NI | Pfizer-BioNTech/ 30 μg, Moderna/3 0 μg () | NI | NI | Both vaccines provide significant protection against SARS-Cov-2 infection, with Pfizer-BioNTech demonstrating 95% efficacy at 30 g recommended for those older than 16 years. In contrast, the COVID-19 effectiveness of the Moderna vaccine, which is recommended for individuals 18 years of age and older at a dose of 30 g, is 94.5 %. | [364] |

*Age designations= according to the American Medical Association; Neonates or newborns (birth to 1 month), Infants (1 month to 1 year), Children (1 year through 12 years), Adolescents or teenagers (13 years through 17 years, Adults (18 years or older) and Older adults (65 and older)

*NI= Not informed

## 7. Future strategies to prevent the spreading of COVID-19

Protective measures are non-pharmaceutical interventions that substantially reduce the rate of SARS-CoV-2 transmission, such as physical separation, mask use, telework, hand cleansing, isolation, and appropriate respiration [365]. Haque et al. [366] reported that protective measures are highly effective in controlling the spread of COVID-19 because no evidence was found to support the claim that any vaccine provides lifetime protection against COVID-19, and achieving zero COVID-19 is not feasible due to the emergence of novel variants that may cause new waves of an ongoing pandemic. In addition, modifying existing vaccines to combat novel SARS-CoV-2 variants is time-consuming [367]. Protective measures can therefore prevent the spread of COVID-19.

The World Health Organization (WHO) exhorted individuals to take precautions such as maintaining a safe distance, wearing tight-fitting masks, cleansing their hands, avoiding congested areas, and getting immunized to reduce their risk of contracting COVID-19 variants [368]. Any failure to implement these preventative measures results in SARS-CoV-2 infection. Depending on the quality and fit of surgical masks, SARS-CoV-2 transmission is reduced by forty to 60 %. While facemasks can reduce the risk of Omicron transmission, they do not offer comprehensive protection against Covid-19 [369,370].

Boretti [371] also stated that wearing masks significantly reduces SARS-CoV-2 transmission and that surgical masks are more effective than fabric masks. Another randomized research by Abaluck et al. [372] revealed that utilizing surgical masks lowered the prevalence ratio of COVID-19 by 11 %. This research also found that older persons (those over 60) had the most significant decline (35 %) in the prevalence ratio for COVID-19. Furthermore, changing facemasks regularly increased efficacy [373]. Kim et al. [374] did a meta-analysis on the efficiency of N95 masks in reducing the transmission of SARS-CoV-2 and found that these masks were 70 % effective compared to surgical masks (30 % effectiveness). N95 breathing masks decrease aerosol exposure by 98.5 %, giving more protection than surgical masks. Additionally, N95 masks protect against nosocomial transmission [368].

Omicron variants have four times the viral load (nasopharyngeal) of other variants [375]; thus, disinfectants decreased the incidence of COVID-19 due to Omicron variants by 77 % [376]. Xu et al. [377] reported 2.44, 2.63, and 3.53 risk ratios for acquiring COVID-19 through inappropriate breathing practices, maintaining social distance, and hand cleansing. Physical separation and remaining at home can also reduce the risk of contracting COVID-19 by 12 % and 51 %, respectively [378].

## 8. Conclusion and future recommendations

SARS-CoV-2 is a highly contagious and pathogenic coronavirus primarily responsible for the COVID-19 pandemic. It poses acute and chronic effects when entering the human body through ACE2 receptors. SARS-CoV-2 chsronic or long-term effects are more detrimental than acute ones because acute diseases are short-lived and chronic conditions are long-lasting; therefore, our study investigated the long-term effects of the virus on various human organs by disrupting the physiological process of pulmonary, cardiovascular, neurological, reproductive, gastrointestinal, musculoskeletal, endocrine and lymphatic systems by affecting homeostasis, inducing oxidative stress, inducing secretion of proinflammatory cytokines, endothelial dysfunction, myocardial contractile dysfunction, inflammation of cardiomyocytes, imbalance in $Ca^{2+}$ flux, inducing thrombosis, inducing chronic neuropathy, spermatogenic dysfunction, reducing antral follicle development, causing cytotoxic injury, inducing myokine storm, dysfunction of corticotroph cells and thyrotropic cells, inducing lymphadenopathy. This comprehensive literature review showed that older adults are more susceptible to the long-term effects of SARS-CoV-2. Conventional and advanced diagnostic strategies for the virus are presented, with their comparisons based on sensitivity, specificity, and cost-effectiveness, and it is demonstrated that implementing COVID-19 diagnosis instruments based on artificial intelligence is the most promising and effective global pandemic strategy. Regarding the treatment of the long-term effect of SARS-CoV-2 in post-COVID-19 patients, pharmacological and non-pharmacological strategies are most effective for getting relief from the long-term impact of COVID-19. Long-term physical exercise may provide innate immune protection for COVID-19 patients by reducing cytokine release syndrome and improving the response of T cells, increasing mobilization of natural killer, and CD8 T cells into the blood and tissues, which ultimately reduces fever, nausea, fatigue, and muscle weakness and pain. Different pharmacological tools, such as sulodexide, ivabradine, apixaban, omega-3 fish oil, atorvastatin, fampridine, vortioxetine, ultra-micronized palmitoylethanolamide, luteolin, ivermectin, tadalafil, and vitamin D supplementation, bismuth subsalicylate help restore the proper function of the various human organ system damaged by the SARS-CoV-2 infection. Vaccination is a promising defense mechanistic tool to boost the immunity of individuals against SARS-CoV-2; that includes inactivated viral vaccines, live attenuated viral vaccines, viral vectored vaccines, protein subunit vaccines, mRNA vaccines, DNA vaccines, nanoparticle vaccines, and these vaccine types significantly reduce the adverse long-term effects of SARS-CoV-2 in patients recovered from COVID-19. Furthermore, the comparative effectiveness of different vaccines may vary, but all types of vaccines significantly reduced SARS-CoV-2 infection. But no vaccine was reported that provides lifetime protection against COVID-19; therefore, protective measures (non-pharmaceutical interventions), such as physical separation, mask use, telework, hand cleansing, isolation, and appropriate respiration significantly reduce the rate of SARS-CoV-2 transmission. As a prospective direction for this study, the following research areas are recommended for further investigation.

1. To investigate the comparative long-term effects of various strains of SARS-CoV-2 in vaccinated individuals of multiple age groups to determine which strain is more pathogenic.

2. ACE2 receptors provide an attachment site for SARS-CoV-2 and allow the virus to infiltrate the human body; consequently, what if modifying the expression of the ACE2 receptor prevents SARS-CoV-2 from entering human cells?

3. Investigate various receptors (aside from ACE2) and entry pathways or routes for viruses into the human body.

4. More research is required to investigate the long-term effects of SARS-CoV-2 in individuals who have recovered from COVID-19 more than three years ago.

5. Concerning the long-term effects of SARS-CoV-2, it is necessary to propose a comparative analysis of multiple and single vaccine types in individuals of varying ages.

6. Investigate the smartphone-based AI algorithm's beneficial and harmful long-term effects for COVID-19 detection.

7. The effects of various exercise combinations, such as resistance training versus strength training, on the organ systems of post-COVID-19 patients require further investigation.

**Acknowledgment**

We thank the support from the National Natural Science Foundation of China 31970752; 32350410397; Science, Technology, Innovation Commission of Shenzhen Municipality JSGG20200225150707332,JCYJ20220530143014032,WDZC20200820173710001, WDZC20200821150704001;Shenzhen Medical Academy of Research and Translation,D2301002;Shenzhen Bay Laboratory Open Funding, SZBL2020090501004;

**Credit Author Statement (Authors' contributions)**

Muhammad Akmal Raheem: Idea, conceptualization, study design, literature search, writing original draft and data curation. Muhammad Ajwad Rahim: review and editing. Ijaz Gul; Idea, software management, and investigation. Md. Reyad-ul-Ferdous: review and editing. Can Yang Zhang: drawing figures and review. Dongmei Yu: review & editing. Vijay Pandey: review and editing. Ke Du: investigation and formal analysis. Runming Wang: formal analysis and data curation. Sanyang Han: data curation. Yuxing Han: writing review and editing. Qin Peiwu: funding acquisition, resources, investigation, project administration, supervision, review, and editing.

**Conflicts of interest**

We have no conflicts of interest associated with this publication. I confirmed that the manuscript has been read and approved for submission by all the named authors.